# Correlative image learning of chemo-mechanics in phase-transforming solids

Haitao D. Deng[1,2], Hongbo Zhao[3], Norman L. Jin[1], Lauren Hughes[4], Benjamin Savitzky[4], Colin Ophus[4], Dimitrios Fraggedakis[3], András Borbély[5], Young-Sang Yu[6], Eder Lomeli[1], Rui Yan[2], Jueyi Liu[2], David A. Shapiro[6], Wei Cai[7], Martin Z. Bazant[3,8], Andrew M. Minor[4,9,*], William C. Chueh[1,10,*]

[1]Department of Materials Science and Engineering, Stanford, Stanford, CA 94305, USA

[2]Institute for Computational and Mathematical Engineering, Stanford, CA 94305, USA

[3]Department of Chemical Engineering, Massachusetts Institute of Technology, 77 Massachusetts Avenue, Cambridge, Massachusetts 02139, USA

[4]National Center for Electron Microscopy, Molecular Foundry, Lawrence Berkeley National Laboratory, Berkeley, CA 94708, USA

[5]Mines Saint-Etienne, Univ. Lyon, CNRS, UMR 5307 LGF, Centre SMS, F - 42023 Saint-Etienne, France

[6]Advanced Light Source, Lawrence Berkeley National Laboratory, Berkeley, CA 94720, USA

[7]Department of Mechanical Engineering, Stanford University, Stanford, CA 94305, USA

[8]Department of Mathematics, Massachusetts Institute of Technology, 77 Massachusetts Avenue, Cambridge, Massachusetts 02139, USA

[9]Department of Materials Science and Engineering, University of California, Berkeley, CA 94720, USA

[10]Stanford Institute for Materials and Energy Sciences, SLAC National Accelerator Laboratory, Menlo Park, CA 94025, USA

*Correspondence to: aminor@berkeley.edu; wchueh@stanford.edu

**Abstract:**

Constitutive laws underlie most physical processes in nature. However, learning such equations in heterogeneous solids (e.g., due to phase separation) is challenging. One such relationship is between composition and eigenstrain, which governs the chemo-mechanical expansion in solids. In this work, we developed a generalizable, physically-constrained image-learning framework to algorithmically learn the chemo-mechanical constitutive law at the nanoscale from correlative four-dimensional scanning transmission electron microscopy and X-ray spectro-ptychography images. We demonstrated this approach on Li$_X$FePO$_4$, a technologically-relevant battery positive electrode material. We uncovered the functional form of composition-eigenstrain relation in this two-phase binary solid across the entire composition range ($0 \leq X \leq 1$), including inside the thermodynamically-unstable miscibility gap. The learned relation directly validates Vegard's law of linear response at the nanoscale. Our physics-constrained data-driven approach directly visualizes the residual strain field (by removing the compositional and coherency strain), which is otherwise impossible to quantify. Heterogeneities in the residual strain arise from misfit dislocations and were independently verified by X-ray diffraction line profile analysis. Our work provides the means to simultaneously quantify chemical expansion, coherency strain and dislocations in battery electrodes, which has implications on rate capabilities and lifetime. Broadly, this work also highlights the potential of integrating correlative microscopy and image learning for extracting material properties and physics.

**Main:**

Learning constitutive laws in solids — material-specific relations that mathematically map multiple physical quantities to one another — could lead to advancements in energy[1], electronics[2] and many other scientific and engineering applications[3–6]. For example, the composition–eigenstrain relation in solids, such as the linear response postulate in Vegard's law[7,8], governs chemo-mechanics[9–12] – that is, how a material changes shape with compositional variation. This widely used empirical relationship in materials science is essential for the prediction of device performance, reliability and degradation[13–16]. For a single-phase, homogeneous material, constitutive relations involving composition is straightforwardly characterized[7]. However, learning such laws for materials which are heterogenous at the nanoscale is substantially more difficult. One example is phase-separating solids[17–20]. Experimentally, the composition–eigenstrain relation cannot be directly characterized within the thermodynamically-unstable miscibility gap[19–21]. First-principles predictions are limited by the computational cost spanning multi-length scales and structural landscapes required to accurately describe phase-separating heterogeneous materials[6,22].

New approaches to extract constitutive relations in solids such as those between composition and eigenstrain are needed. Over the past few decades, significant progress has been made in high-resolution X-ray, electron, and neutron microscopy, enabling direct visualization of the nanoscale morphology, structure, chemistry and strain[23–28]. The spatially correlated pixels inside microscopy images inherently embed the constitutive relations of interest, thus offering a new pathway to address the aforementioned challenge[29–32]. To date, experimental images have been used for estimating physical parameters[15,31,33,34] and identifying limiting processes[25,32,35] by focusing on a small number of features (such as pattern wavelengths, interface thicknesses, etc).

Using data-driven approaches, progress has been made on distilling atomic-scale images into physical insights[32,36], but much less so at the meso- and macro-scale. Accurate constitutive-relation learning through microscopy-image inversion requires the integration of experiments, data analysis, and theory to enforce connection to natural physics[32,37,38].

In this work, we algorithmically learn the nanoscale composition–eigenstrain relation through partial differential equation (PDE)-constrained optimization[39,40] in $Li_XFePO_4$. This well-characterized and widely adopted battery positive electrode[19,25,41–43] has a large miscibility gap, with X spanning between approximately 0.05 and 0.9[20,41]. Despite more than two decades of progress, two fundamental questions remain for $Li_XFePO_4$: (1) what is the extent of elastic coherency (commonly assumed in models[15,44]) and plastic deformation at phase boundaries, and (2) how does the metastable solid solution expand and contract inside the miscibility gap? To quantify the Li composition and lattice strain in thick-particle specimens, we employed X-ray spectro-ptychography[23,45,46] and 4D scanning transmission electron microscopy[24,47] (4D-STEM), respectively. Using an image-learning framework employing physical constraint and regularization[29,40], we uncovered a mostly linear Li composition–eigenstrain relation (i.e., a constant chemical expansion coefficient) in the two-phase binary system across the entire composition range, experimentally validating Vegard's law at the nano-scale. Benchmarked against the non-physically-constrained direct regression results, our image-learning approach reduced the model error by 19%. Importantly, through our framework, we also quantified nanoscale heterogeneities in the residual strain field (i.e., those beyond the compositional eigenstrain and coherency elastic strain), which we attributed to plastic deformation through dislocation density analysis. This result was verified independently through X-ray diffraction line profile analysis[48]. These findings are foundational to improving capacity retention and rate capabilities of phase-transforming battery electrodes[14,49], and the method is generalizable to most crystalline materials. Our approach integrates microscopy, materials science, and image learning, highlighting the benefits of accurately extracting physical laws from scientific imaging data.

The pipeline for inverse image learning from correlative images is shown in Fig. 1. The procedure consists of three main steps: (1) correlative microscopy[50], (2) image registration, and (3) PDE-constrained optimization. Our objective is to learn the Li composition–eigenstrain relation, $f(X_{Li})$, in $Li_XFePO_4$ for all compositions ($0 \leq X \leq 1$). Therefore, nanoscale measurement of the composition and strain, particularly at the phase boundaries, is necessary. We employed correlative 4D-STEM and X-ray spectro-ptychography to probe the local lattice constant and Li composition in 10 platelet particles ($2 \times 4$ µm wide and ~300 nm thick) across 54,508 experimental image pixels[50]. Briefly, 4D-STEM acquires 2D convergent beam electron diffraction patterns at every pixel of a 2D STEM raster scan; X-ray spectro-ptychography employs a phase retrieval algorithm to reconstruct the real image as a function of X-ray energy using coherent diffraction patterns whose fields of illumination are overlapped on the sample. The spatial resolution for 4D-STEM (National Center for Electron Microscopy TitanX) reaches 1 nm[24,47] while X-ray ptychography (Advanced Light Source beamline 7.0.1.2 (COSMIC)) achieves sub-10 nm[23]. We note that STEM electron energy loss spectroscopy, a commonly-used approach, is not possible here because of the specimen thickness (< 100 nm is generally required), as is often the case for technologically-relevant materials such as those for batteries, etc. To ensure the formation of phase-separation boundaries in individual platelet particles, we controlled the Li composition by chemical delithiation rather than by electrochemical cycling to avoid interparticle "mosaic" phase separation (see **Methods** and Fig. S1 for details and general characterization)[51].

For 4D-STEM, taking the unit cell of LiFePO$_4$ as our unstrained reference, we extracted the lattice parameters $a$ and $c$, shear strain $\gamma$, and rotation angle $\theta$ from the nanoscale diffraction patterns at each real-space position (Fig. 1, step 1a). The zone axis is along the crystallographic b axis. The obtained lattice information was then converted into normal strains along the a and c axis. The details of the process are in **Methods** and Fig. S2. For X-ray spectro-ptychography, as shown in Fig. 1, step 1b, the local Li composition $X_{Li}$ was inferred from the nanoscale X-ray absorption spectra at the Fe L$_3$ edge (see **Methods** and Fig. S3) by assuming local electroneutrality. We acknowledge that the electroneutrality will be violated near interfaces and dislocations. Nonetheless, the very short Debye screening length (< 0.1 nm) in Li$_X$FePO$_4$ likely means that such non-electroneutral domains are below our imaging resolution.

In the second step, we align the images by maximizing the correlation between the lattice parameters and Li composition (details in **Methods**). The aligned image stack (Fig. 1, step 2) contains both compositional and crystallographic features, namely $\{X_{Li}, a, c, \gamma, \theta\}$. The constitutive law governing the chemo-mechanics is implicitly embedded in the pixels along the feature space dimension. Finally, in the third step, we learn $f$ while obeying the stress equilibrium condition:

$$\mathcal{F} := \nabla \cdot \sigma = 0 \qquad (1)$$

where the left-hand side is the divergence of the stress tensor. In other words, this inverse image learning problem is a PDE-constrained optimization problem, as illustrated in Fig. 1, step 3.

Fig. 2a shows the aligned Li composition and lattice parameter maps of a typical phase-separated Li$_X$FePO$_4$ particle. Our technique provides correlative measurement of composition and lattice constant across phase boundaries. The characteristic widths match within error between ptychography (123±39 nm, Fig. 2b, Fig. S12) and 4D-STEM (132±41 nm, Fig. 2c, Fig. S12) imaging, consistent with some previous literature[25,52]. While others have reported sharp interfaces (~ 10 nm in width)[53], we argue that a sharp phase boundary is unlikely in our case as it would otherwise suggest an energetically unfavorable >20 degree tilting angle relative to the zone axis[54]. Furthermore, the phase-field model sensitivity on the interface width is explored in Fig. S13. Before performing inverse image learning of chemo-mechanics, we first carried out a standard data-driven analysis: pixel-wise lattice constant vs. Li composition correlation is plotted across 10 particles (Fig. 2d-f). We classified the $Li_XFePO_4$ particles into three categories based on their particle-averaged Li composition, $0 \leq \overline{X} \leq 0.15$ ("FP", Fig. 2d), $0.15 < \overline{X} < 0.85$ ("$L_XFP$", Fig. 2e), and $0.85 \leq \overline{X} \leq 1$ ("LFP", Fig. 2f). In LFP and FP, no phase boundaries were observed, and the mean lattice parameters depend weakly on Li composition. For $L_XFP$, while the mean lattice parameters near the end-member compositions were consistent with those of LFP and FP, there were two major differences: (1) the mean lattice parameter changes as a function of composition and (2) the variance of the lattice parameter inside the miscibility gap (~16% of all pixels, mostly coming from phase boundaries, Fig. S12) given a fixed composition was larger than that of LFP and FP.

As mentioned earlier, rigorously distilling the Li composition–eigenstrain relation from the correlated data requires an accounting of various strains under the constraint of stress equilibrium. Specifically, the experimentally observed nonuniform lattice strain arises from three sources: composition-dependent eigenstrain, coherency strain and dislocations[55]. We began by first neglecting the dislocation strain in our forward model; we term this the coherency strain model "M1", consistent with existing literature[15,44]. The interpretation of the model error is in **Supplementary Information**. Validity of this assumption will be examined through residual strain analysis later. The remaining two strains sum to the total strain $\epsilon$ measured by 4D-STEM;

stress is thus expressed as the inner product of stiffness $C$ (known) and elastic strain: $\sigma = C:(\epsilon - f(X_{Li}))$, constrained by the stress-equilibrium PDE (labeled as $\mathcal{F}$). For discretization during inversion, we directly computed the 2D displacement field $d$ from $\{a, c, \theta, \omega\}$ (see **Supplementary Information**). Consequently, the inverse image learning problem is: given $d$ and $X_{Li}$, what is the composition-dependent eigenstrain relation $f$ under constraint $\mathcal{F}$?

To find $f$, we parameterized the relation in the following form:

$$f_i = \sum_{n=1}^{N} a_{ni} L_n(X_{Li}), \quad i = 1, 2 \tag{2}$$

where $i = 1, 2$ represents the strain in the a and c directions, respectively; $L_n$ are the Legendre polynomials defined over $X_{Li} \in [0,1]$; and $a_{ni}$ are the regression coefficients. To the first order, Eqn. 2 gives Vegard's Law commonly adopted in alloys[7]. Regularization was implemented by fine-tuning the cutoff number N, promoting the sparsity of the fitting parameters to obtain simple physics (details in **Supplementary Information**). Our optimization objective was therefore to minimize the difference between the experimental displacement and model prediction $\hat{d}$ based on the PDE constraint $\mathcal{F}$:

$$\min J(d, \hat{d}) := \|d - \hat{d}\|_M^2, \quad s.t. \; \mathcal{F}(d, f) = 0 \tag{3}$$

where $M$ is the difference measure and taken as the Euclidean $L_2$ norm. The explicit expressions and derivations are provided in **Supplementary Information**. For benchmarking, we also fitted $f$ without considering elastic strain or stress equilibrium (hereinafter, "direct regression model M0").

The training and test results for model M1 are shown in Fig. S5 d–i. We found that N = 2 best avoids systematic overfit. Fig. 2c shows the recovered compositional eigenstrain relation for the a and c directions, with its 95% point-wise confidence band highlighted in the shaded regions.

This result indicates that the eigenstrain varies almost linearly with composition, providing a direct experimental validation of Vegard's law at the nanoscale. In contrast, the benchmark (model M0, direct regression without stress equilibrium constraint, same regularization with 4 parameters) yielded a more non-linear compositional eigenstrain behavior, as shown by the dashed lines in Fig. 2c, which is similar to previous report on nanosized $Li_XFePO_4$[19] (derived from X-ray diffraction measurements without spatial resolution). The difference in the compositional eigenstrain relation between models M0 and M1 is particularly acute in the chemical expansion coefficient (i.e., derivative of eigenstrain w.r.t. Li composition), averaging ~28% root mean square difference across the full composition range as shown in Fig. 2d). We note that the chemical expansion coefficient is essential for constructing the chemical potential in phase-field theories[15,44] (derivation in **Supplementary Information**). Physically, the difference between the two models is the coherency (elastic) strain, which model M0 neglects. As we detail later, physics-constrained inverse image learning produces less error than the direct regression model despite having fewer parameters. This approach highlights the importance of physics-informed image inversion rather than a pure data-driven approach, the latter of which does not guarantee consistency with solid mechanics.

Having extracted the Li composition–eigenstrain relation, we generate additional chemo-mechanical insights from the residual strain field. As a case study, we focus on a single phase-separated particle with an average Li composition of $\bar{X}$ = 0.51. The Li composition map and the total strain field (measured by 4D-STEM) are presented in Fig. 3a. Given the experimental Li composition and $f$, coherency strain is computed through 2D phase field simulations (Fig. S7, details in **Supplementary Information**). Next, we visualize the residual strain field, defined as the total strain minus the compositional eigenstrain and coherency strain. Fig. 3b shows significant local heterogeneities in the residual strain, which are consistently observed throughout $L_XFP$ particles (Fig. S9). Such local hotspots impact strain compatibility (details in **Supplementary Information)**.

The root mean square residual strain incompatibility vs. particle-averaged Li composition reaches a maximum at $\bar{X} \sim 0.5$ (Fig. 3d). These strain hotspots may be due to morphological defects; the particles studied are not uniformly thick and contain voids, as observed from X-ray tomography (Fig. S4). To assess this hypothesis, we employed a 3D phase-field simulation on similarly-sized and shaped particles with internal pores (Fig. S8). However, no local hotspots in the residual strain were observed. Having ruled out pore effects, the residual strain heterogeneities likely arise from misfit dislocations. We note that such dislocations are not necessarily localized at the phase boundaries as the boundaries move during delithiation (Fig. 3c). Additionally, we note that while TEM is the preferred method to visualize dislocations, the large specimen thickness (~300 nm) made imaging impossible due to dynamic scattering[56].

Indeed, simulation based on misfit dislocations among potential dislocation systems generated residual strain maps similar to the experimental observation (Fig. 3c), except for regions where 4D-STEM measurement has higher uncertainty (Fig. S10). Motivated by these results, we constructed a dislocation model (M2) that includes compositional eigenstrain (same parameterization as M1), coherency strain and dislocation effects (parameterized by dislocation density of different dislocation types), detailed in **Supplementary Information**. Fig. S6 shows that the dislocation model error was minimized when the Burgers vector was along the crystallographic c-direction, with an average density of ~$282\mu m^{-2}$ (Fig. 3c). We highlight that this dislocation strain field is otherwise impossible to quantify without accounting for compositional and coherency strains[57].

We assess the effectiveness of the physical constraint for inverse image learning through residual error analysis. Fig. 3e compares the model errors of direct regression model (M0), coherency strain model (M1), and dislocation model (M2). Benchmarked against M0 (baseline), M1 reduces the model error by 19%, and M2 reduces the model error by an additional 24%. From a data science perspective, the principle of parsimony favors M1 over M0 because of its increase in model accuracy yet decrease in the number of parameters. More broadly, this result emphasizes the benefits of having appropriate physical constraints for inverse image learning. On the other hand, independent validation of the M2 is necessary as the local dislocation density is a spatially dependent variable that scales quadratically with the image size.

We turned to X-ray powder diffraction to independently verify the dislocation density. Specifically, we employed a variance-based line profile analysis[48] to evaluate the dislocation densities and types. As shown in Fig. 4, the dislocation-induced broadening was the highest for FP, intermediate for $L_{0.5}FP$ (powder with an average Li composition of 0.5), and smallest for pristine LFP (which has never seen the formation of phase boundaries). Furthermore, because each dislocation carries a unique angle-dependent broadening signature, we used non-negative least square minimization to identify the major dislocation systems. Table 1 shows the most likely dislocation combinations for LFP, $L_{0.5}FP$, and FP. Relevant analysis is presented in **Supplementary Information**. The total dislocation density increases with the delithiation extent. Specifically, the dislocation system 'D', representing dislocations with [001] Burgers vector and slip plane (100) with an estimated density of 306 ± 97 $\mu m^{-2}$ is consistent with the dislocation density determined from model M2. Other candidates (whose line directions are not aligned with [010]) do not significantly alter the local strain field in the crystallographic ac-plane measured through 4D-STEM. The result is also consistent with the fact that the formation of misfit dislocations with smaller Burgers vector magnitude is thermodynamically more favorable. Interestingly, after complete delithiation, the dislocation contributing to the ac-plane strain disappears, consistent with residual strain analysis (Fig. 3d). Investigation of dislocation dynamics is underway. Finally, the estimated dislocation density is of similar orders of magnitude to that of other battery positive electrodes observed during/after cycling[27,42,58].

In summary, we presented an inverse image learning framework that combines state-of-the-art high-resolution X-ray and electron microscopy with PDE-constrained optimization. The functional form of the composition–eigenstrain relation across the full Li composition range was quantified in the phase-separating system $Li_XFePO_4$. Benchmarked against the constraint-free baseline model, our approach reduced the model error by 19%. Removing the compositional and coherency strain also revealed the residual strain heterogeneities in phase-separated particles. Local heterogeneities are likely due to edge dislocations of type [001] (100), which further reduced the model error by an additional 24%. Previous studies have reported a correlation between dislocations and battery rate capabilities as well as lifetime;[27,42,58] our findings provide the means and motivation for attaining a more fundamental understanding of dislocations inside battery electrodes. Broadly speaking, our approach to infer physics through image inversion is highly applicable to other systems, particularly those that traditionally face thermodynamic or characterization limitations[17,59]. We anticipate that the same approach can be adapted for dynamic systems, where the laws governing the reaction kinetics and transport can be learned[25,29]. The proposed method demonstrates the potential of merging correlative microscopy with physics-constrained image learning for scientific discovery.

## Methods:

The methods are included in Supplementary Information.

## Data availability:

The 4D-STEM and X-ray microscopy data associated with this manuscript can be found at https://data.matr.io/6/.

## Acknowledgements:

This work was supported by the Toyota Research Institute through the Accelerated Materials Design and Discovery Program. X-ray ptychography development was supported by DOE, Office of Basic Energy Sciences, Division of Materials Sciences and Engineering (contract DE-AC02-76SF00515). This research used resources of the Advanced Light Source, which is a DOE Office of Science User Facility under contract no. DE-AC02-05CH11231. Work at the Molecular Foundry was supported by the Office of Science, Office of Basic Energy Sciences, of the U.S. Department of Energy under Contract No. DE-AC02-05CH11231. Use of the Stanford Synchrotron Radiation Lightsource, SLAC National Accelerator Laboratory, is supported by the U.S. Department of Energy, Office of Science, Office of Basic Energy Sciences under Contract No. DE-AC02-76SF00515. Part of this work was performed at the Stanford Nano Shared Facilities (SNSF)/Stanford Nano-fabrication Facility (SNF), supported by the National Science Foundation under award ECCS-1542152. We thank Chirranjeevi Gopal, Patrick Herring, Abraham Anapolsky for assistance in 4D-STEM data pipeline setup. We thank Neel Nadkarni for insightful discussions on the mechanics that inspired this work. We thank Mehrdad Kiani for insightful discussions on dislocations. We thank Halwest Mohammad, Yinyu Ye for helpful discussions on PDE-constrained optimization algorithms. We thank Harry Thaman, and Emma Kaeli for manuscript review.

**Contributions:**

H.D.D., N. J., W.C.C., and A.M. conceived the experiments. H.D.D., N.J., and E.L. performed the synthesis and materials characterization. H.D.D. and N.J. performed the STXM and ptychography experiments. H.D.D. performed the STXM and X-ray spectro-ptychography data analysis. Y.-S.Y. and D.A.S. assisted in the STXM and ptychography experiments. L.H. performed the 4D-STEM experiments. C.O. performed the image registration. L.H. and B.S. performed the 4D-STEM analysis. H.D.D., H.Z., and M.Z.B. developed and performed the inverse image learning optimization. R. Y., and J. L. assisted in the early algorithmic exploration of PDE-constrained optimization. H.D.D. and W.C. performed the 2D phase-field simulation and dislocation-density optimization. D.F. performed the 3D phase-field simulation. H.D.D. performed the residual strain analysis. H.D.D., W.C., and A.B. performed the X-ray line profile analysis. Y.-S.Y. analyzed the ptycho-tomography data. H.D.D. prepared the manuscript. All authors contributed to the discussion of the results and writing of the manuscript.

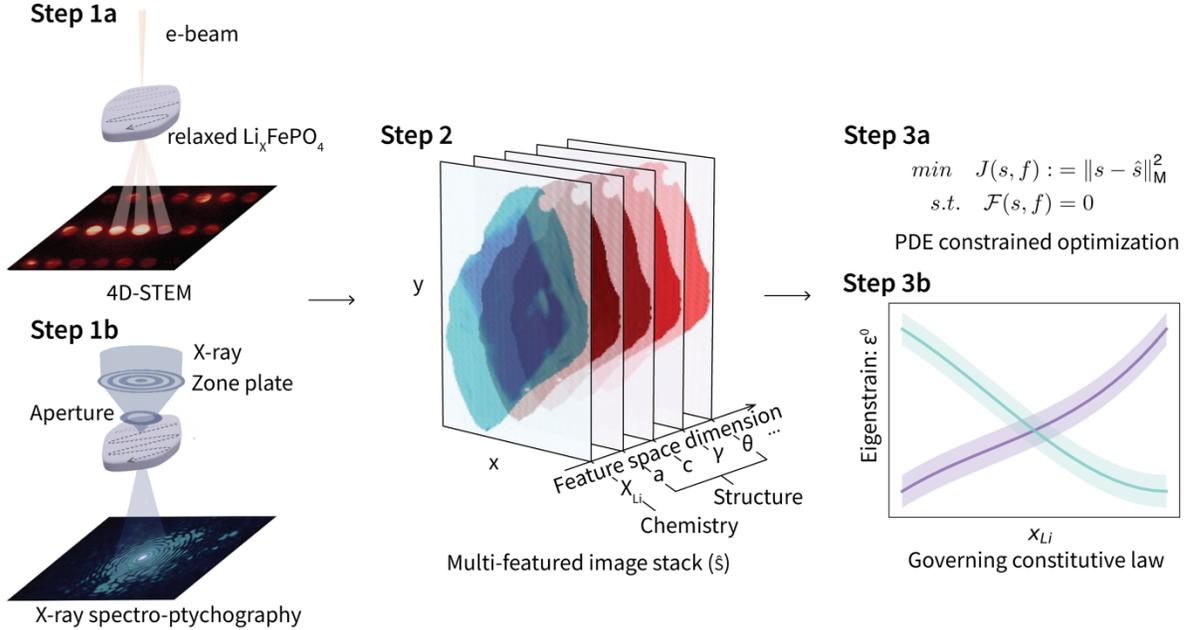

**Fig.1|Schematic of inverse image learning framework for constitutive equations.** The learning procedure is divided into 3 steps. Step 1, correlative microscopy characterization of the relaxed Li$_X$FePO$_4$ (for the same particle). Specifically, four-dimensional scanning transmission microscopy (4D-STEM, 1a) and X-ray spectro-ptychography (1b) were employed to probe the local structure and Li composition of chemo-mechanically relaxed Li$_X$FePO$_4$ with a resolution down to 16 nm. Step 2, a multi-featured image stack $\hat{s}$ (containing composition $X_{Li}$, lattice parameter $a, c$, shear $\gamma$ and rotation angle $\theta$) is obtained after image registration. The pixels in $\hat{s}$ embed the constitutive law $\epsilon^0 = f(X_{Li})$, describing the coupling of Li composition and mechanics. Step 3, inverse learning from $\hat{s}$. The inverse problem is solved via a partial differential equation (PDE)-constrained optimization (3a). The optimization takes the input from the data as well as the physical constraint of mechanical equilibrium into consideration. Subsequently, the governing constitutive chemo-mechanical law is uncovered (3b).

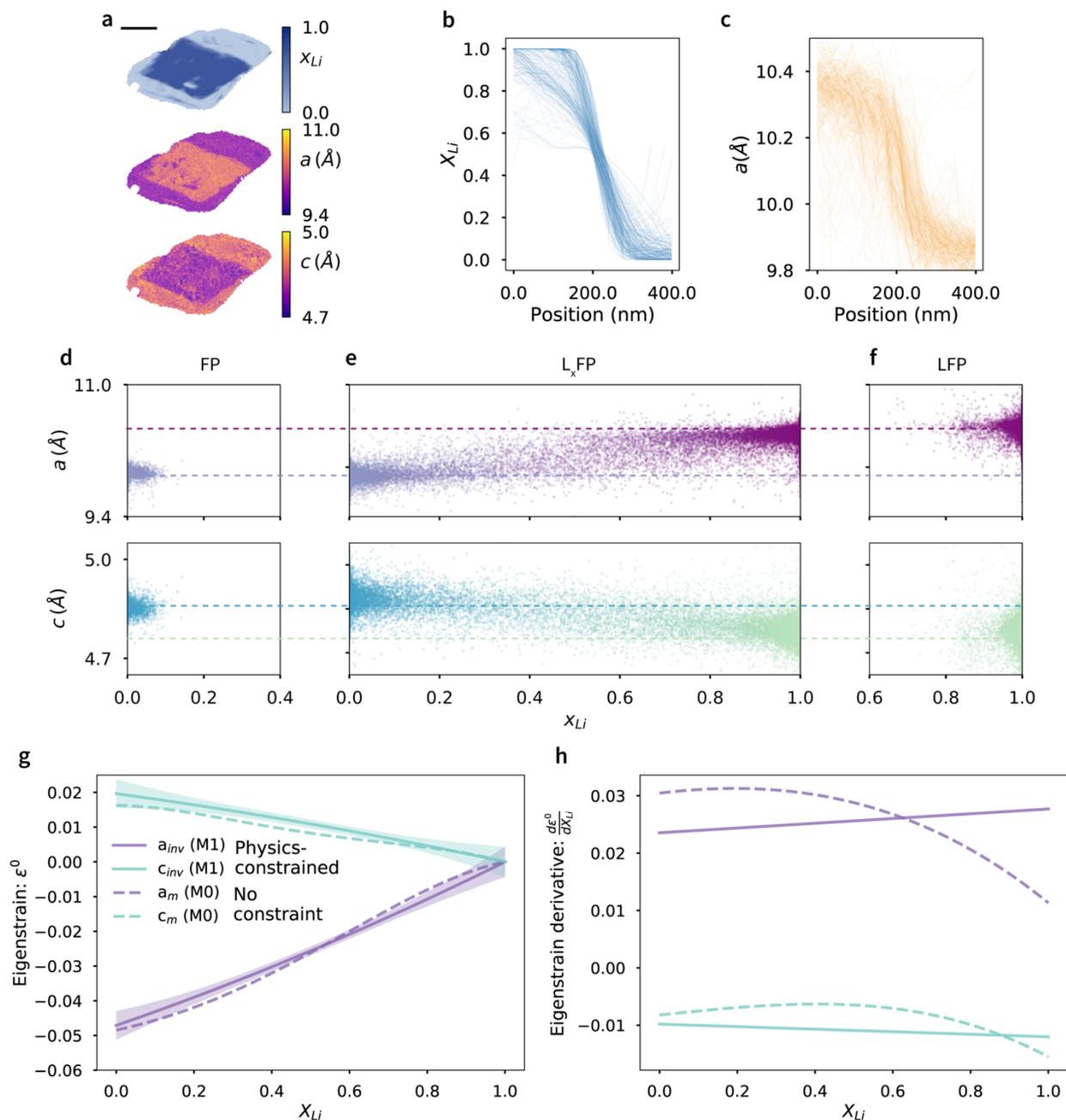

**Fig.2|Inverse image learning of composition–eigenstrain relation. a**, Correlative Li composition (i.e., X in $Li_XFePO_4$) and lattice parameter map of chemo-mechanically relaxed biphasic $Li_{0.51}FePO_4$. A clear phase boundary is observed in the images. The scale bar is 500 nm. **b,c**, 1D Li composition and lattice parameter line profile across the interface. Each line represents a line cut perpendicular to the phase boundary, detailed in Fig. S12. The characteristic widths match within error between composition (123±39 nm) and lattice parameter (132±41 nm). **d-f**, Statistical pixel-wise lattice–composition correlation diagrams in $Li_XFePO_4$ of all particles measured. Correlations of in-plane a, c

lattice parameters and Li composition are shown for "FP" ($0 \leq \bar{X} \leq 0.15$, left), phase-separated "L$_x$FP" ($0.15 < \bar{X} < 0.85$, middle) and "LFP" ($0.85 \leq \bar{X} \leq 1$, right) respectively. The average lattice parameter in the phase-separated particles at a given composition lies between that of its end members while the spread is larger inside the miscibility gap, due to mechanical strain. **g**, Constitutive compositional eigenstrain (unitless) inversely learned from all phase-separated samples (model M1, purple and green solid lines for a ($a_{inv}$) and c ($c_{inv}$) direction respectively). LiFePO$_4$ is chosen as the unstrained reference state. The point-wise 95% confidence band is in shaded regions. As a comparison, direct regression (model M0, baseline) fits without stress equilibrium constraint are shown as dashed lines ($a_m$, $c_m$). **h**, Comparison of the constitutive eigenstrain relation between M0 and M1. The major difference between the two can be seen in the first derivative of eigenstrain (unitless) w.r.t. Li composition (i.e., chemical expansion coefficient), in that the curvature of the relations obtained is significantly different. The inversely learned curve has an almost constant derivative while M0 result is nonlinear. Our M1 result therefore directly validates Vegard's law of linear response at the nanoscale.

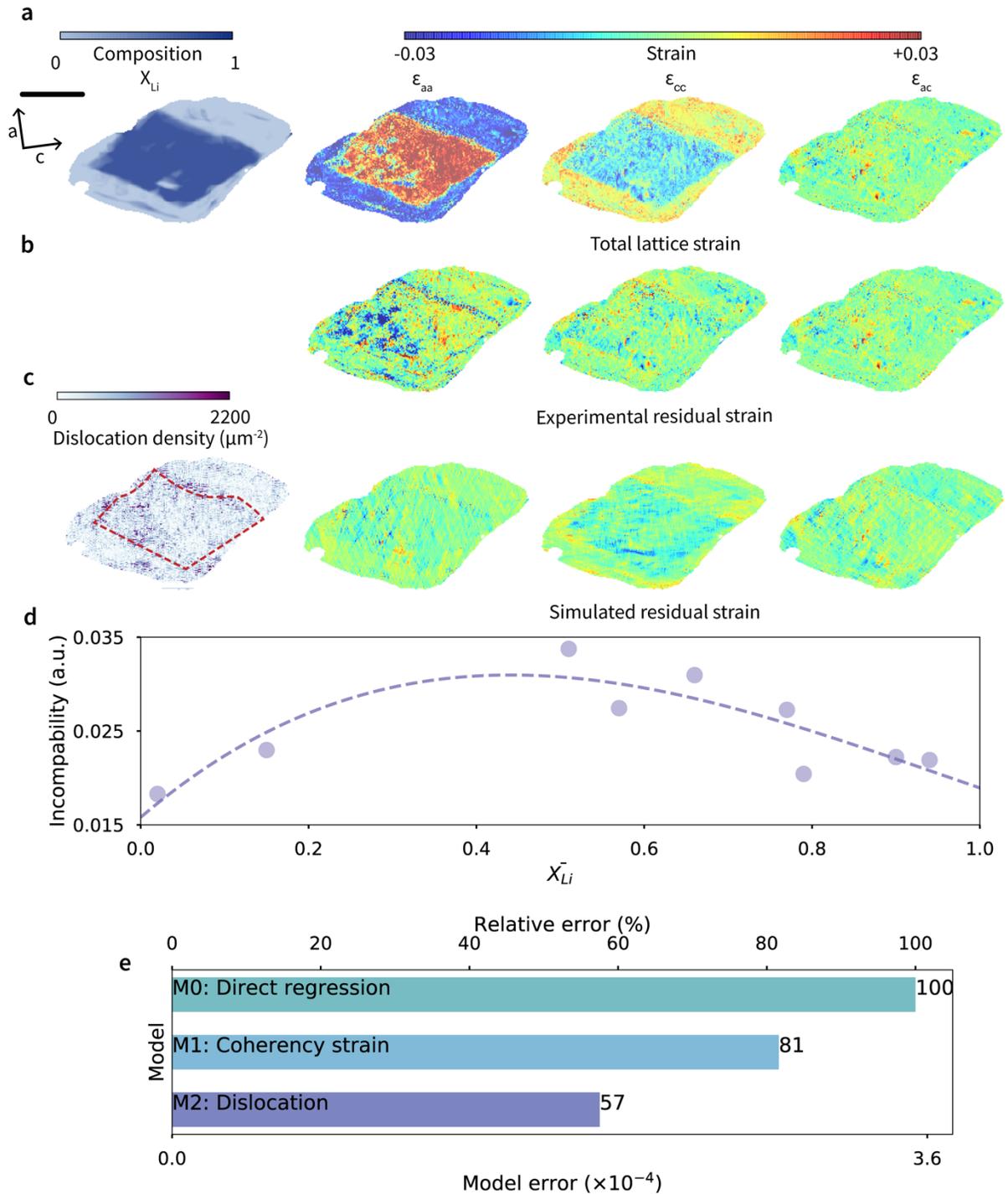

**Fig.3|Chemo-mechanical insights of Li$_X$FePO$_4$. a**, Experimental observation of Li composition and total (lattice) strain (measured by 4D-STEM, unitless) field of Li$_{0.51}$FePO$_4$. Well-defined phase boundaries are formed inside chemo-mechanically equilibrated Li$_X$FePO$_4$, with orientation close to 101 direction. **b**, The residual strain field, defined as the total strain minus the compositional eigenstrain and coherency (elastic) strain. Coherency strain is computed by 2D phase field simulations. The residual strain field is

highly non-uniform and displays local hotspots in $\varepsilon_{aa}$, $\varepsilon_{cc}$ and $\varepsilon_{ac}$. These hotspots are located around the phase boundaries and inside the phase separated regions, indicating the existence of other types of strains, for example dislocations. **c**, Simulated dislocation density of type 'D' [001](100) and its residual strain field. The particle-averaged dislocation density is ~282µm$^{-2}$, consistent with the independent X-ray line profile analysis. The dislocations are not necessarily located at phase boundaries (red line), as interfaces move during delithiation. Many of the local hotspots in the residual strain field are recovered, except for regions where 4D-STEM measurement has higher uncertainty, shown in Fig. S10. **d**, Residual strain incompatibility as a function of particle-averaged Li composition. Effects of the heterogeneities can be quantified through the residual strain incompatibility – that is, the root mean square residual of the compatibility condition at each pixel. The dashed line is a guide to the eye. Visualization of the residual strain field for all particles is in Fig. S9. **e**, Model error– that is, the mean Frobenius norm of residual strain – from the three models: direct regression (M0, baseline), coherency strain (M1) and dislocation model (M2) respectively. For a fair comparison, all model errors were converted into strain. M1 has a 19% reduction in model error from M0 while M2 has a 43% reduction compared with M0. Scale bar: 1µm. Arrows represent the average crystallographic axis.

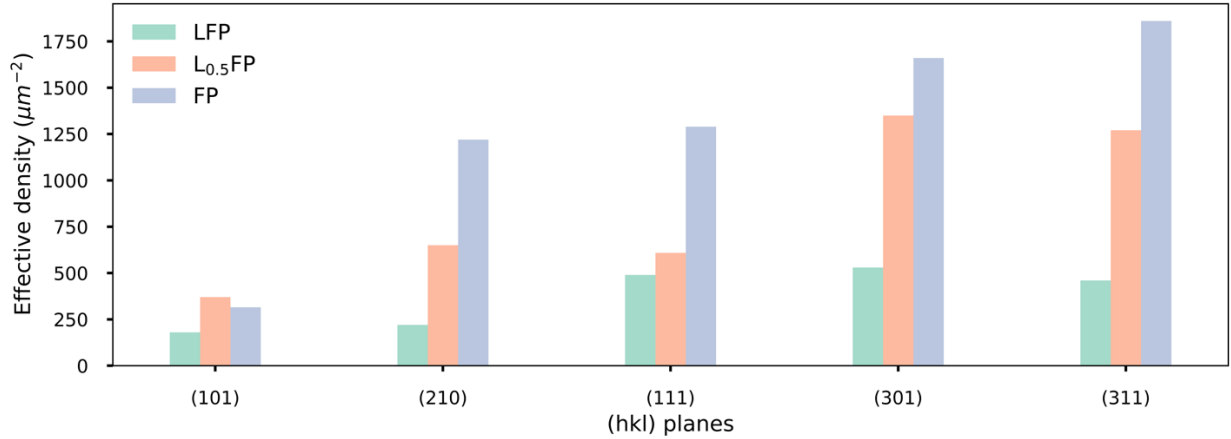

**Fig.4|Dislocation induced X-ray line broadening.** Effective dislocation density as a function of diffraction peaks for LFP, $L_{0.5}FP$ and FP powder samples respectively. The total effective dislocation density increases with the delithiation extent.

**Table 1. Identification of dislocation systems.** Labeled systems are considered for regression. n is the index plane normal, b is the Burgers vector, l is the sensing vector. All candidates are either edge or mixed dislocations. The value in the parenthesis represents the standard deviation. Total dislocation densities increase with the delithiation extent. In the dislocation types with nonnegative values, only type 'D' contributes to the residual ac-plane strain field measurable by 4D-STEM.

| n | Label | b | l | Type | LFP | $L_{0.5}FP$ | FP |
|---|---|---|---|---|---|---|---|
| | | Total density (um$^{-2}$) | | | 1257(80) | 2147(173) | 4619(351) |
| | | Maximum data explained % | | | 87 | 91 | 85 |
| (100) | A | [010] | [001] | Edge | | 590(183) | 1298(183) |
| | B | [010] | [011] | Mixed | | | |
| | C | [010] | [021] | Mixed | | | |
| | D | [001] | [010] | Edge | 192(39) | 306(97) | |
| | E | ½[011] | [01$\bar{1}$] | Mixed | | | |
| | F | ½[011] | [010] | Mixed | | | |
| (010) | G | [001] | [100] | Edge | | | |
| | H | ½[101] | [$\bar{1}$01] | Mixed | 247(16) | 735(106) | 1382(146) |
| (001) | I | [010] | [100] | Edge | 692(67) | 457(149) | |
| | J | [010] | [110] | Mixed | 45(15) | 556(82) | 1939(261) |
| | K | [010] | [120] | Mixed | | | |
| | L | ½[110] | [1$\bar{1}$0] | Mixed | | | |
| | M | ½[110] | [010] | Mixed | | | |
| (0$\bar{1}$1) | N | ½[011] | [100] | Edge | | | |
| ($\bar{1}$10) | O | ½[110] | [001] | Edge | | | |

# Correlative image learning of chemo-mechanics in phase-transforming solids

## SUPPLEMENTARY INFORMATION

1. List of figures
   

2. Supplementary Text
   

3. References

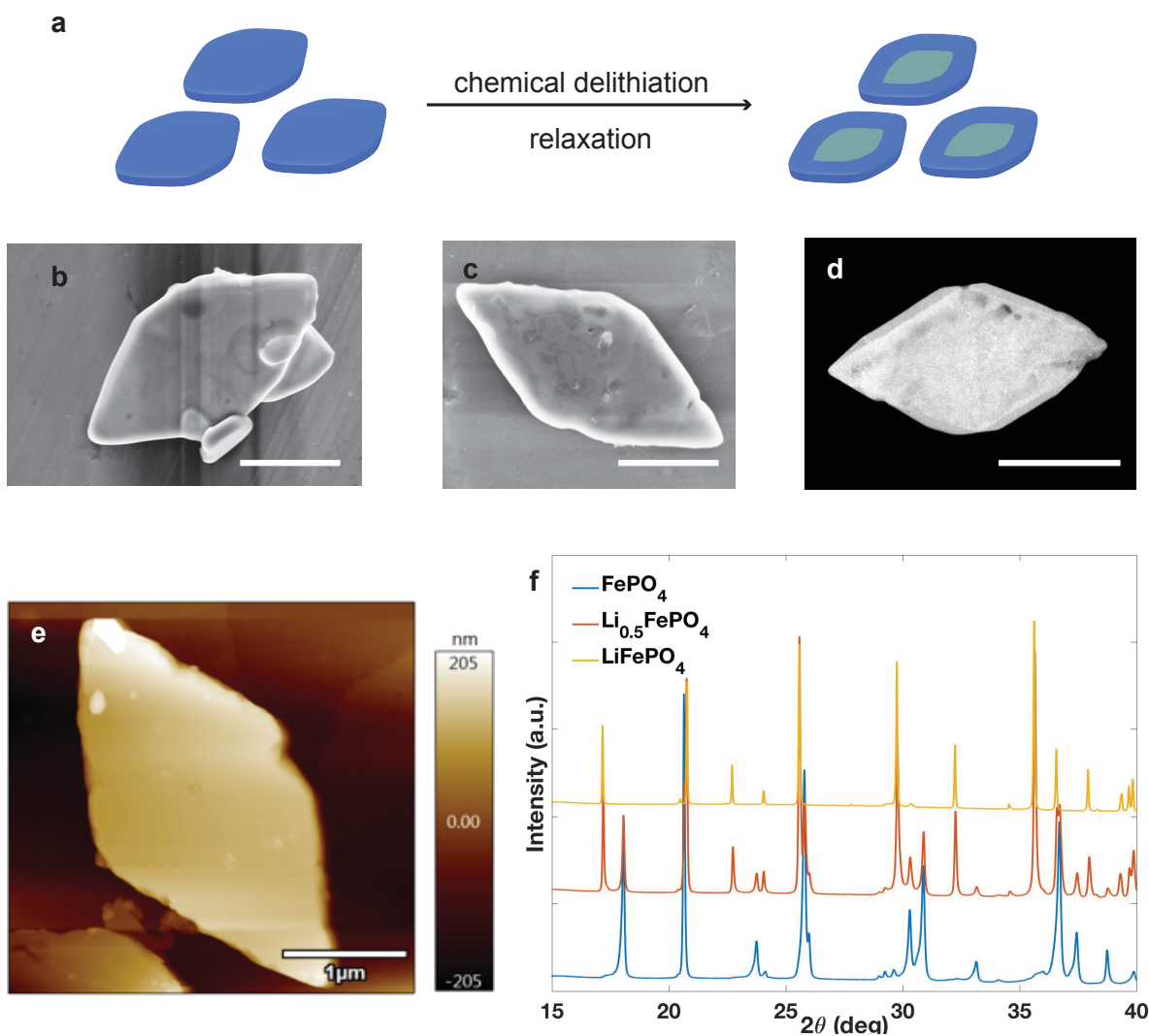

**Figure S5. Chemical delithiation and materials characterization. a**, Schematic of chemical delithiation. **b**, SEM micrograph of diamond-shaped LFP platelet particle. **c**, SEM micrograph of chemically delithiated and relaxed $Li_{0.5}FePO_4$. Comparison of (b) and (c) indicates minimal morphological changes after delithiation. **d**, High-angle annular dark-field image of LFP. **e**, AFM image of an LFP platelet with dimensions of $4 \times 0.3 \times 2\ \mu m^3$. f, Powder XRD patterns of LFP, $Li_{0.5}FePO_4$, and FP. For b–d, the scale bar is 1 $\mu m$.

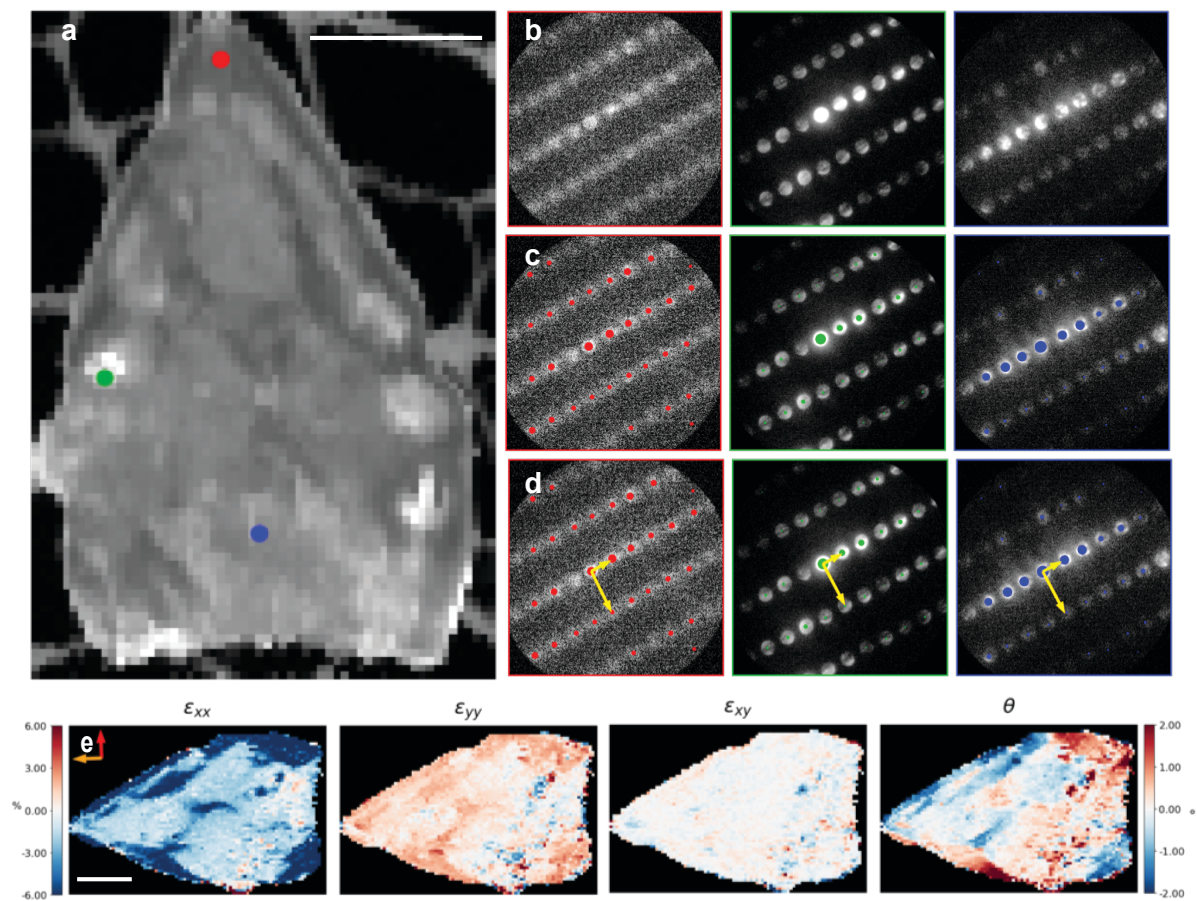

**Figure S2. 4D-STEM and strain analysis. a**, Virtual annular dark field image. **b**–**d**, Extraction of strain from 4D-STEM images. The electron diffraction pattern was first recorded at each position (b), the Bragg disk was then identified (c), and finally, the Bragg disks were fitted to extract the local lattice strain and shear (d). **e**, Normal and shear strain and rotation field of the particle. The zone axis is in the crystallographic b-direction. The crystallographic axes a and c are indicated by the red and orange arrows, respectively, in (e). The scale bar is 1000 nm.

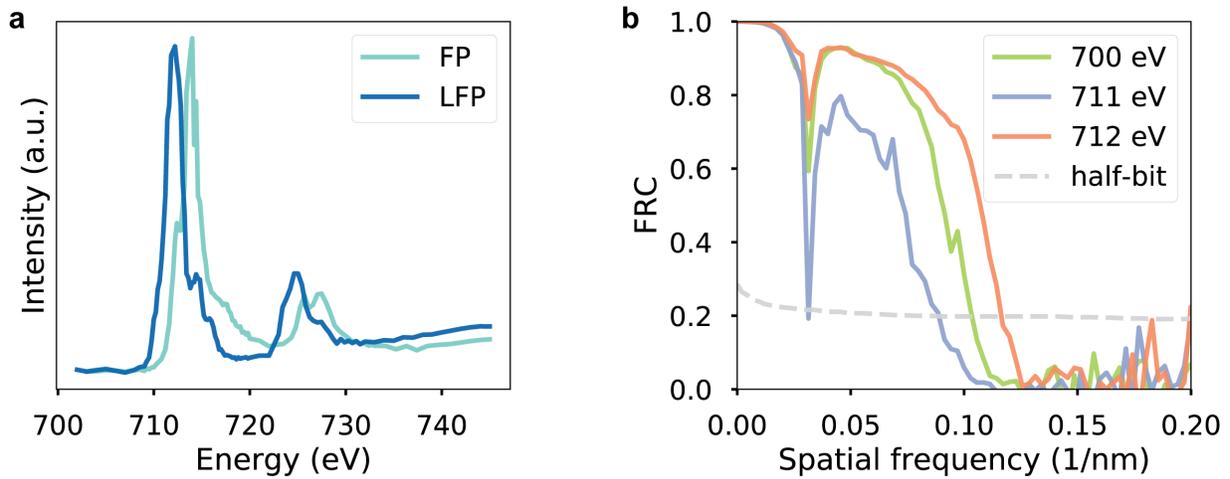

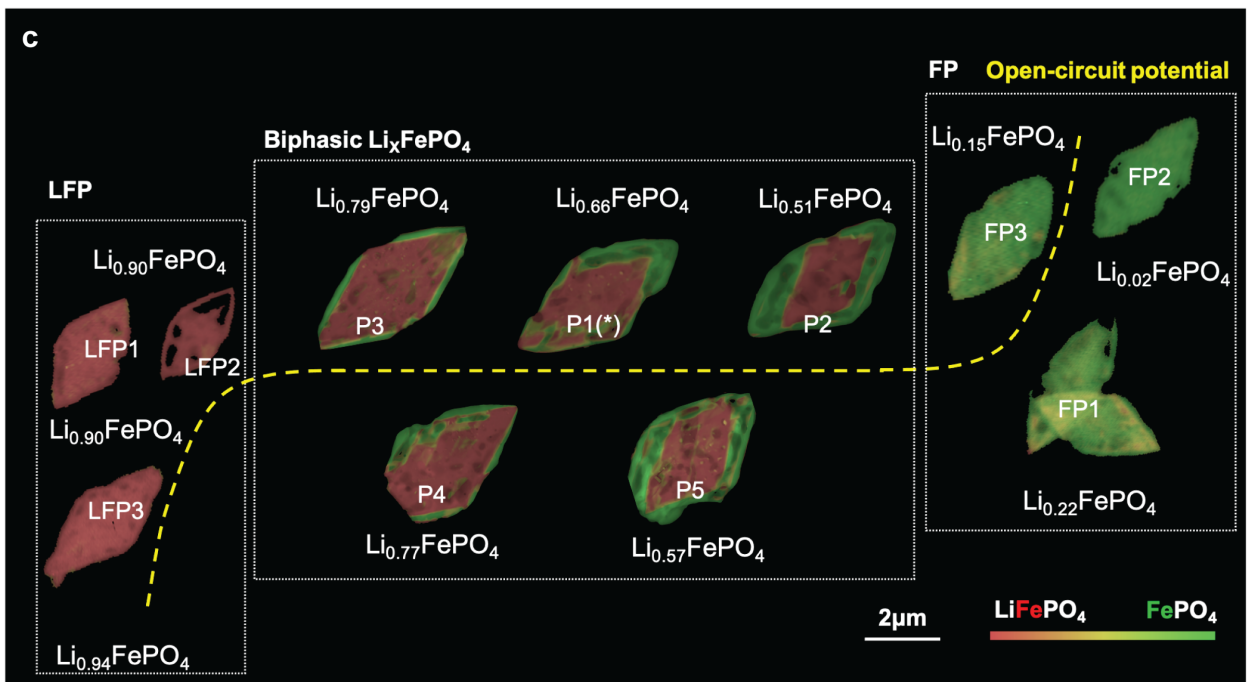

**Figure S3. X-ray spectro-ptychography and STXM. a**, Fe L2 and L3 edge reference spectra for composition determination. **b**, Foureir Ring Correlation (FRC) for $Li_xFePO_4$ (P5). The half-bit spatial resolution is approximately 10 nm for phase reconstruction for ptychography. **c**, All the images collected for STXM and ptychography experiment. For inverse learning, biphasic particles were selected for training and testing. P2–P5 are the training images, and P1 is the test image. The scale bar is 2 $\mu m$.

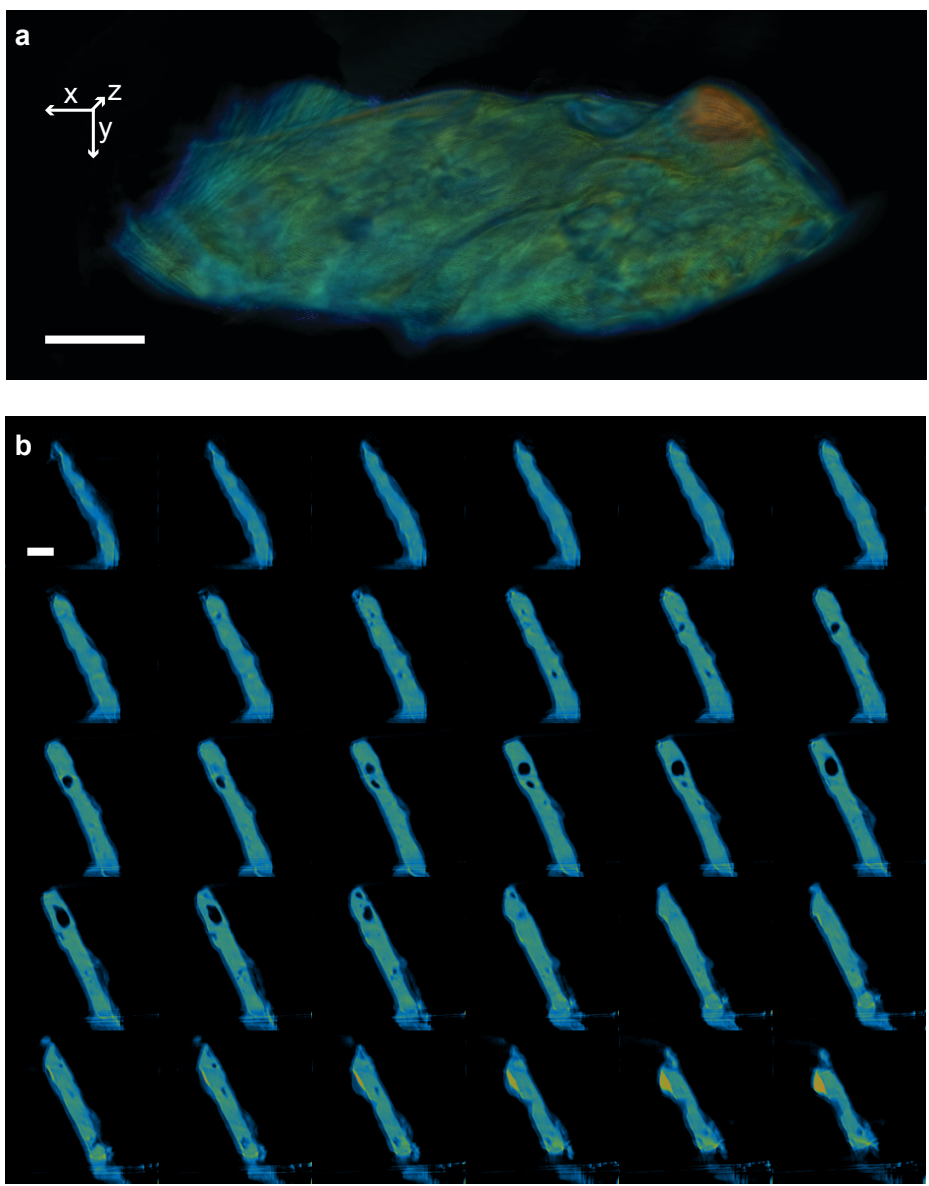

**Figure S4. Characterization of pre-exiting voids using X-ray ptycho-tomography. a**, X-ray ptycho-tomographic micrograph of Li$_X$FePO$_4$. **b**, Cross sections of xy plane in (a), showing that the particle is not uniformly dense and contains voids.

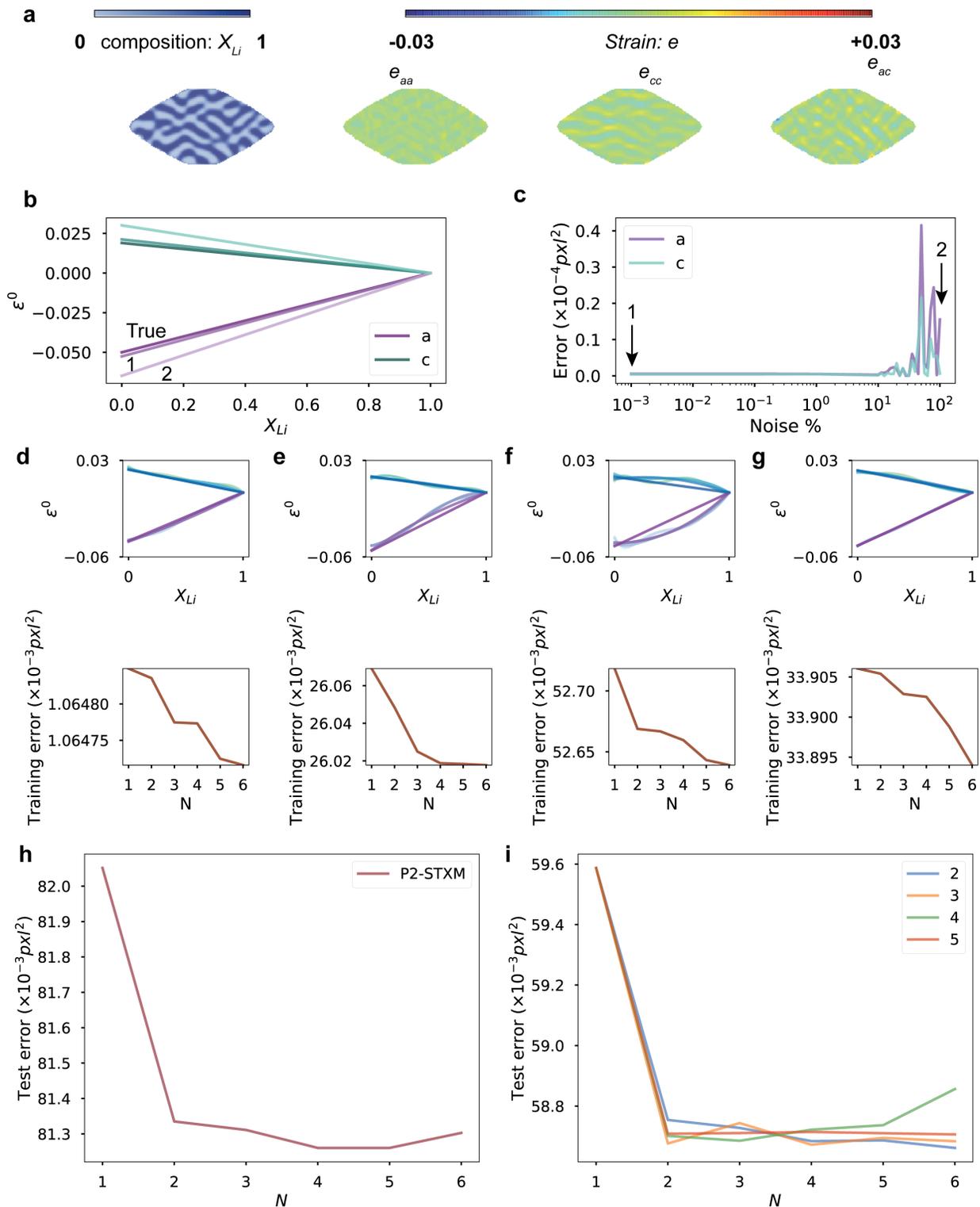

**Figure S5. Inverse learning from simulated data and experimental data. a**, Phase-field simulation of $Li_{0.5}FePO_4$ platelet. The phase-field simulation result was used as the inverse learning input to evaluate the performance. **b**, **c**, Inverse learning performance. We tested the robustness of the methodology by adding different degrees of noise to the input data. The true constitutive laws and the learned ones are plotted in (b), and the difference squared integrated over the entire range is shown in (c). Our method demonstrates that the recovered constitutive law is fairly robust until up to 10% noise in the data. **d**–**g**, Regressed constitutive laws through inverse learning from experimental images and the training error as a function of the number of Legendre polynomial terms $N$. The training error is the objective function (L2 norm squared loss), defined in the supplementary text; pxl represents pixel length. The recovered relation is also dependent on $N$. **h**, **i**, Regularization of parameters in the constitutive-law learning process. **h**, Validation of the model from P2 using STXM image data. **i**, Validation of models trained using P2–P5 tested on P1 image data; the indices are the particle numbers. $N$ is the cutoff number. A cutoff number of 2 preserved some of the non-linear features with the minimum test error.

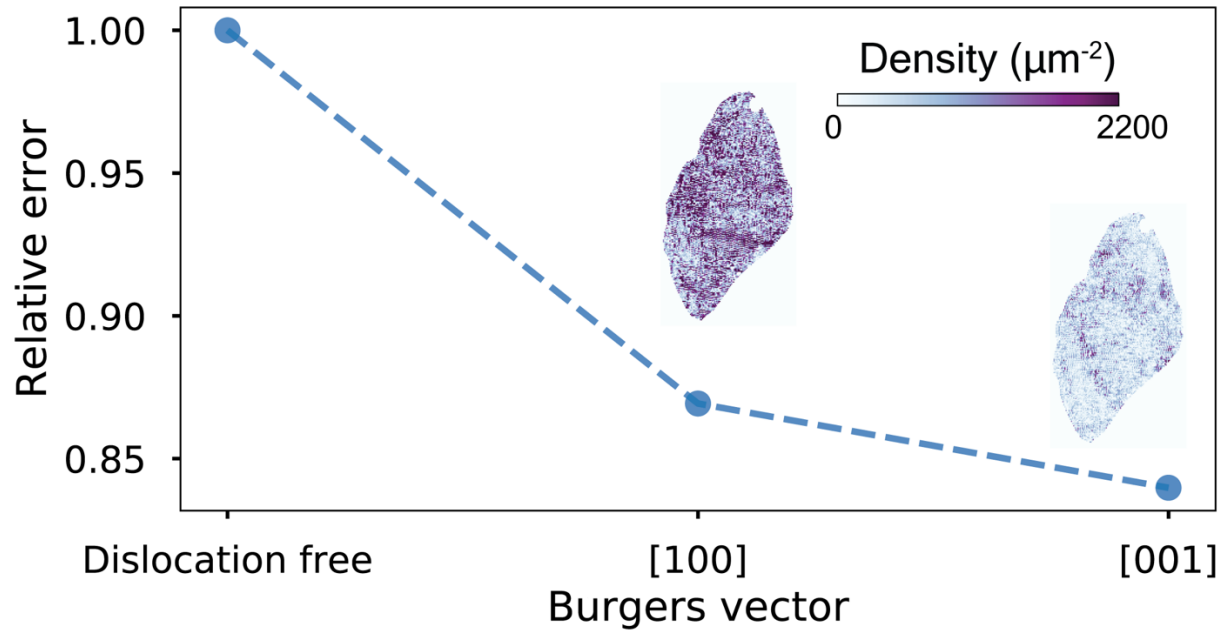

**Figure S6. Dislocation-density optimization.** Potential dislocations with line direction along b-direction were selected and compared. The dislocation with [001] Burgers vector and (100) slip plane has reduced error compared with a dislocation-free model. The optimized particle-average dislocation density is ~282/$\mu m^2$, consistent with the X-ray line analysis. The inset shows the fitted local dislocation density on the particle. The reference for the model error is M1 (coherency strain model).

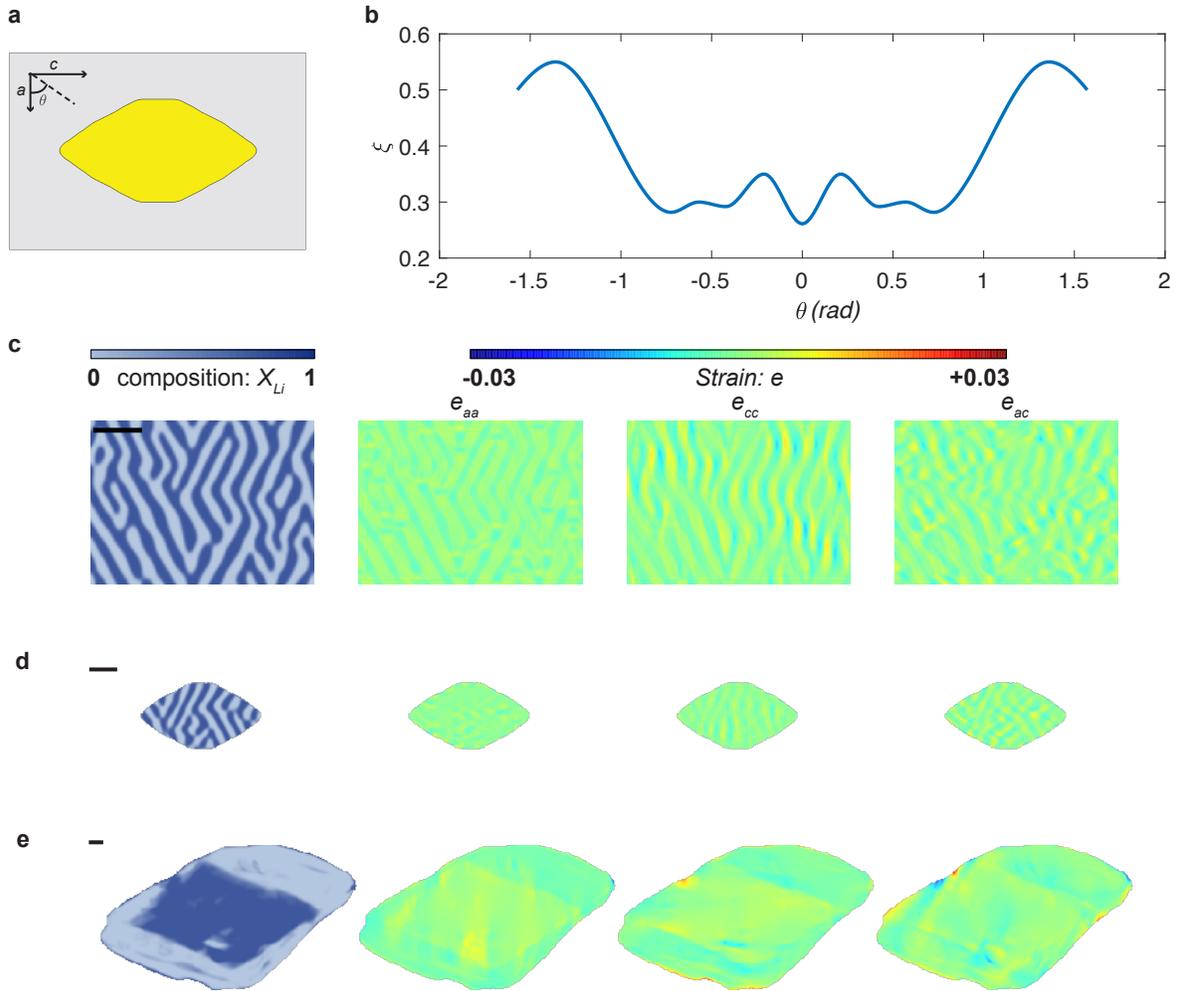

**Figure S7. 2D phase-field simulation. a**, Phase-field simulation yielding an idealized diamond platelet shape by optimizing $\phi$. **b**, Anisotropy in surface energy that generates the diamond platelet shape after relaxation, $\theta$ (in rad) is defined in (a). **c**, Simulated elastic strain field in an infinitely sized crystal with coherent interfaces. **d**, Simulated elastic strain field in a diamond platelet crystal with coherent interface. From (c) and (d), the phase boundary in the bulk orientations are governed by the misfit strain anisotropy, regardless of the boundary position. In addition, negligible elastic strain is generated across the compositional interface, in contrast to interfaces with dislocations, where local dipole-like strain hotspots can be identified. **e**, Coherency strain field of $Li_{0.51}FePO_4$ from calculated given experimental Li composition map and composition–eigenstrain relation obtained from inversion. The scale bar is 200 nm.

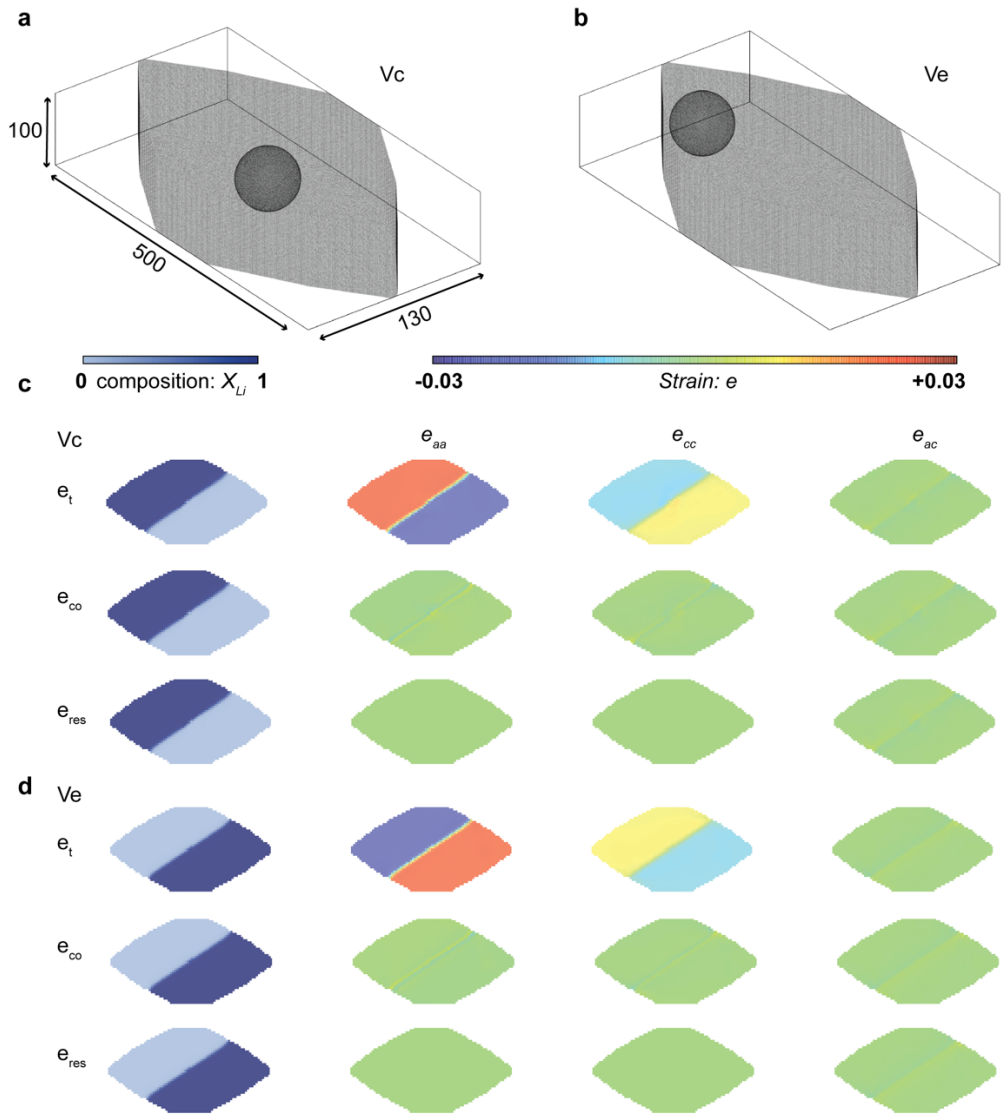

**Figure S8. 3D phase-field simulation**. **a**, Void at the center of the particle. **b**, Void near the edge of the particle. The in-plane direction is along the crystallographic b-axis. **c**, **d**, b-direction averaged concentration and strain fields, including total strain ($e_t$), coherency strain ($e_{co}$), and residual strain ($e_{res}$) using 3D phase-field simulation for cases shown in (a) and (b), respectively. The maximum coherency (elastic) strain at the interface or near the void is <0.6% in the b-direction depth-averaged sense, insignificant at the scale-bar level. The residual strain is highly uniform and close to zero. The dimensions of the particle are 500 nm × 130 nm × 100 nm.

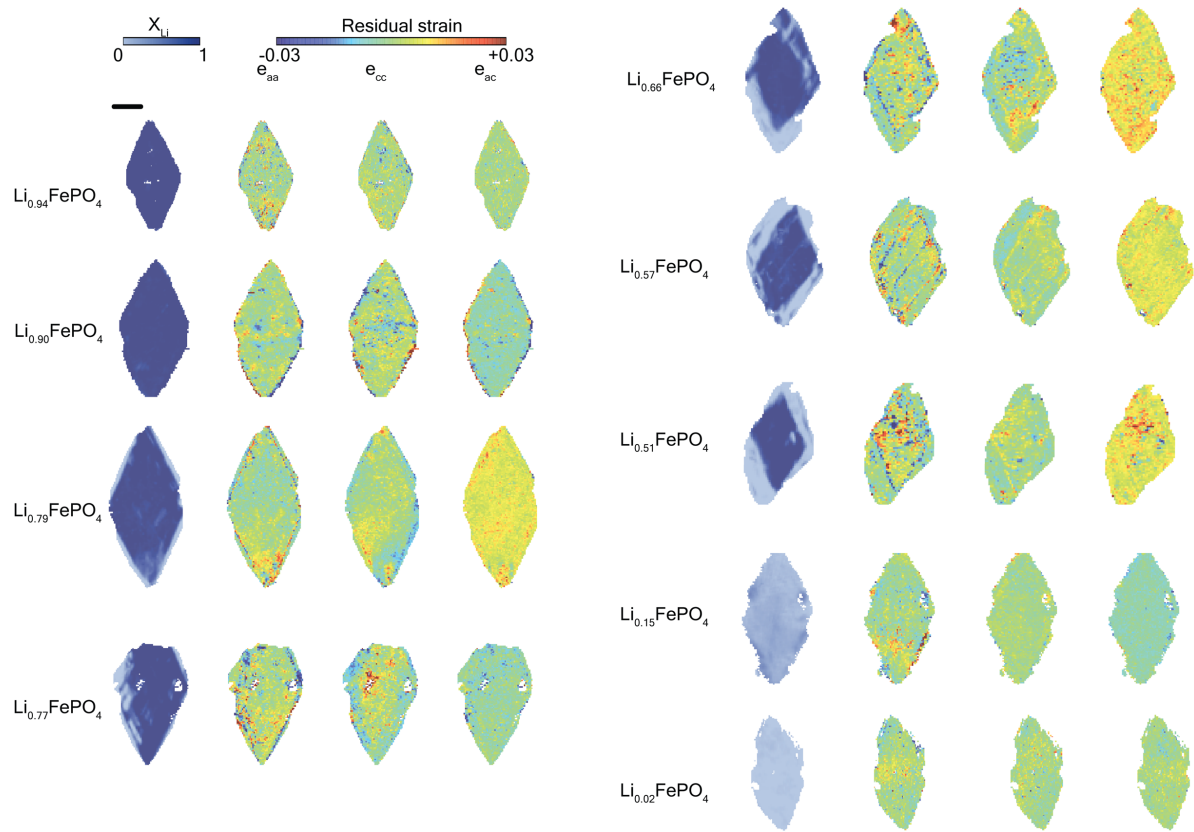

**Figure S9. Residual strain field of Li$_X$FePO$_4$.** The residual strain field for particles with different particle-averaged Li composition. FP1 and LFP2 from Fig. S3 are not analyzed because of overlapping features from other particles. Visually, LxFP has the highest amount of heterogeneities. In addition, we quantified the heterogeneity effect using the root mean square residual strain incompatibility, as shown in Fig. 3d. Details of the residual strain analysis are provided in the supplementary text. The scale bar is 1 μm.

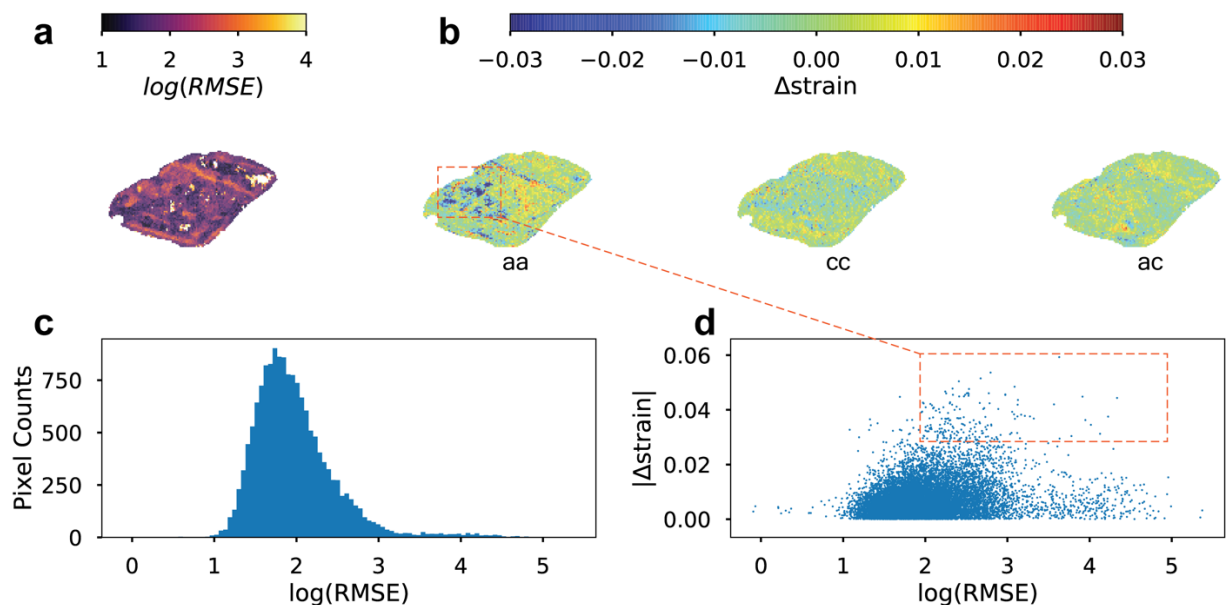

**Figure S10. 4D-STEM strain measurement uncertainty and dislocation model error analysis. a**, Estimated uncertainty in the lattice parameter, defined as the root-mean-square error (RMSE) of the lattice vector fits by leaving out half of the diffraction peaks for uncertainty quantification. A higher value (brighter color) indicates higher uncertainty. The details of the uncertainty quantification for 4D-STEM are discussed elsewhere[1]. **b**, M2 model error Δstrain, namely the experimental residual strain field (Fig. 2b) minus the predicted residual strain field (Fig. 2c). **c**, RMSE error for 4D-STEM measurement. **d**, Correlation between RMSE and Δstrain. Regions with high model error appear to have higher RMSE.

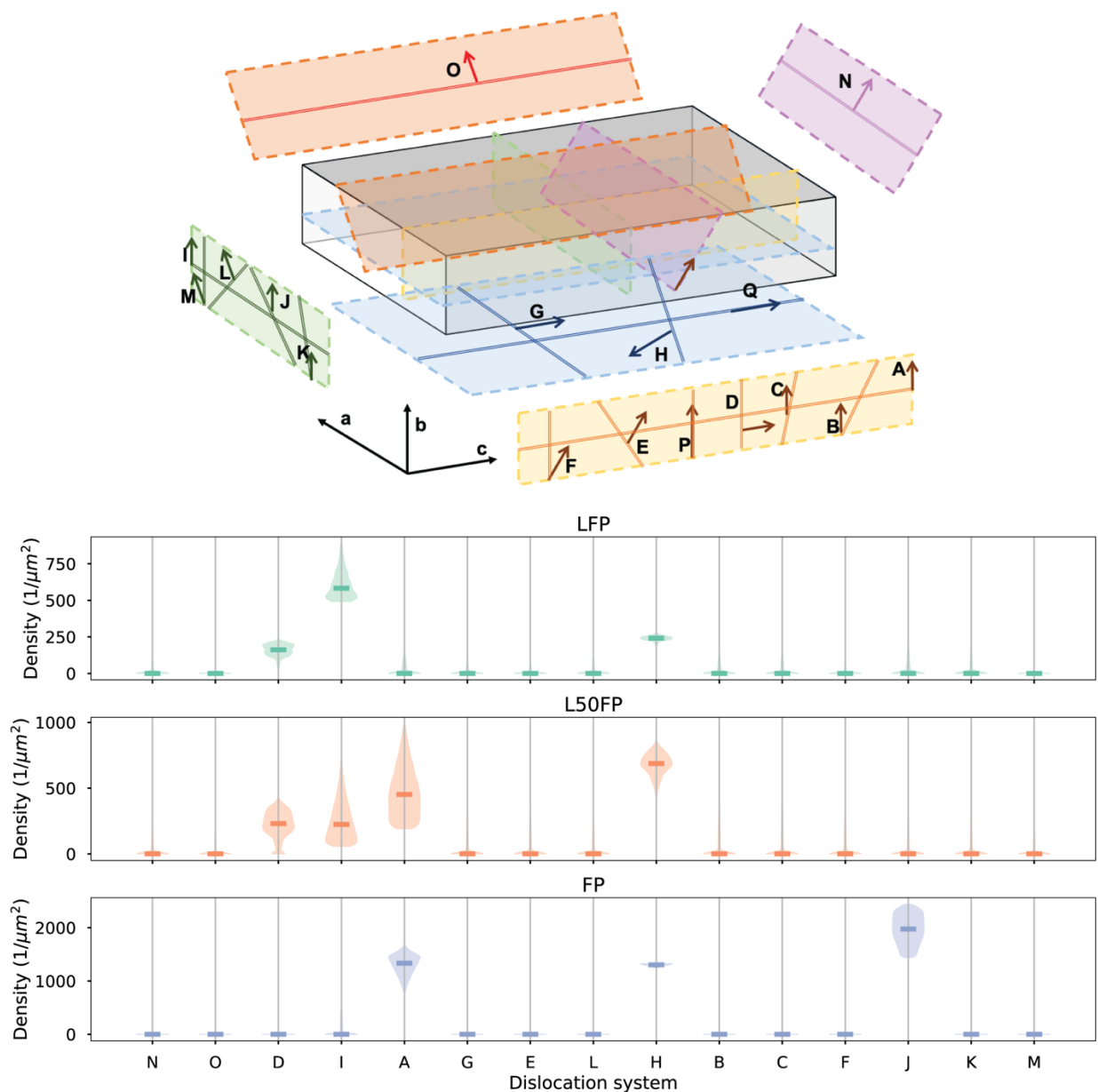

**Figure S11. Dislocation schematics and probability distribution of acceptable dislocation densities.** Dislocation systems considered for XRD line broadening analysis are shown in the top panel, where the arrows are aligned with the Burgers vector, the lines are aligned with the sensing vector, and the planes represent the slip planes. Violin plots are presented in the bottom panel, showing the marginal distribution of each individual dislocation system, given the joint distribution of solutions that achieves a 15% total error tolerance. The horizontal broadening of the shape represents the probability density, with the bar indicating the mean.

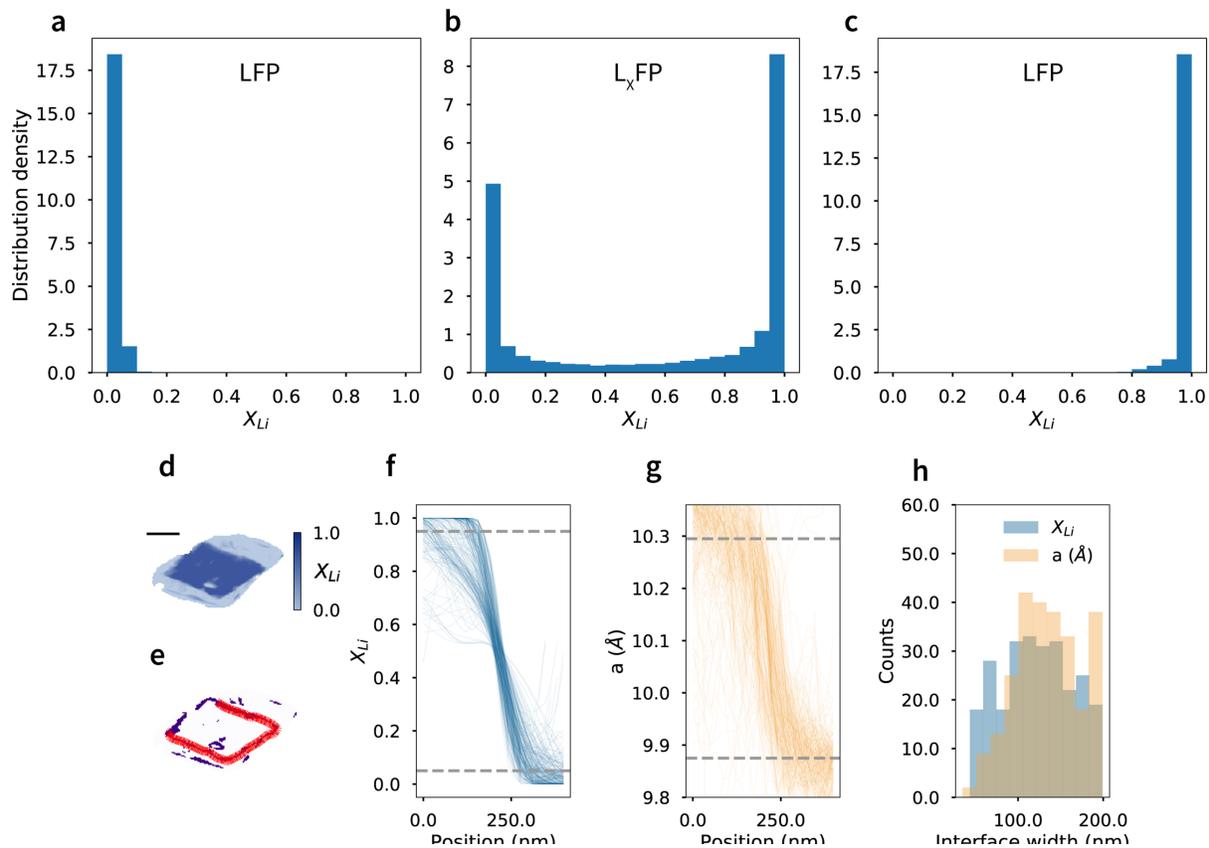

**Figure S12. Li composition distribution and interface width. a-c**, The statistical distribution density of pixel-wise Li compositions for LFP, L$_x$FP and FP respectively. **d**, Li composition map of Li$_{0.51}$FePO$_4$. The scale bar is 500 nm. **e**, Regions in d where Li composition is between 0.15 and 0.85, accounting for ~16% of all pixels. Red lines represent linecuts across the phase boundary. **f,g**, 1D Li composition (f) and lattice parameter (g) profile across the interface. Each line is taken from the linecut (red) shown in e. For Li composition (f), the interface region is defined as 0.05 ≤ X$_{Li}$ ≤ 0.95, while for lattice parameter (g), the interface region is 9.875 Å ≤ a ≤ 10.295 Å, show in grey. h, The statistical distribution of interface width: 123±39 nm (f) and 132±41 nm (g). As can be seen, the interface width distribution is consistent between Li composition and lattice parameter.

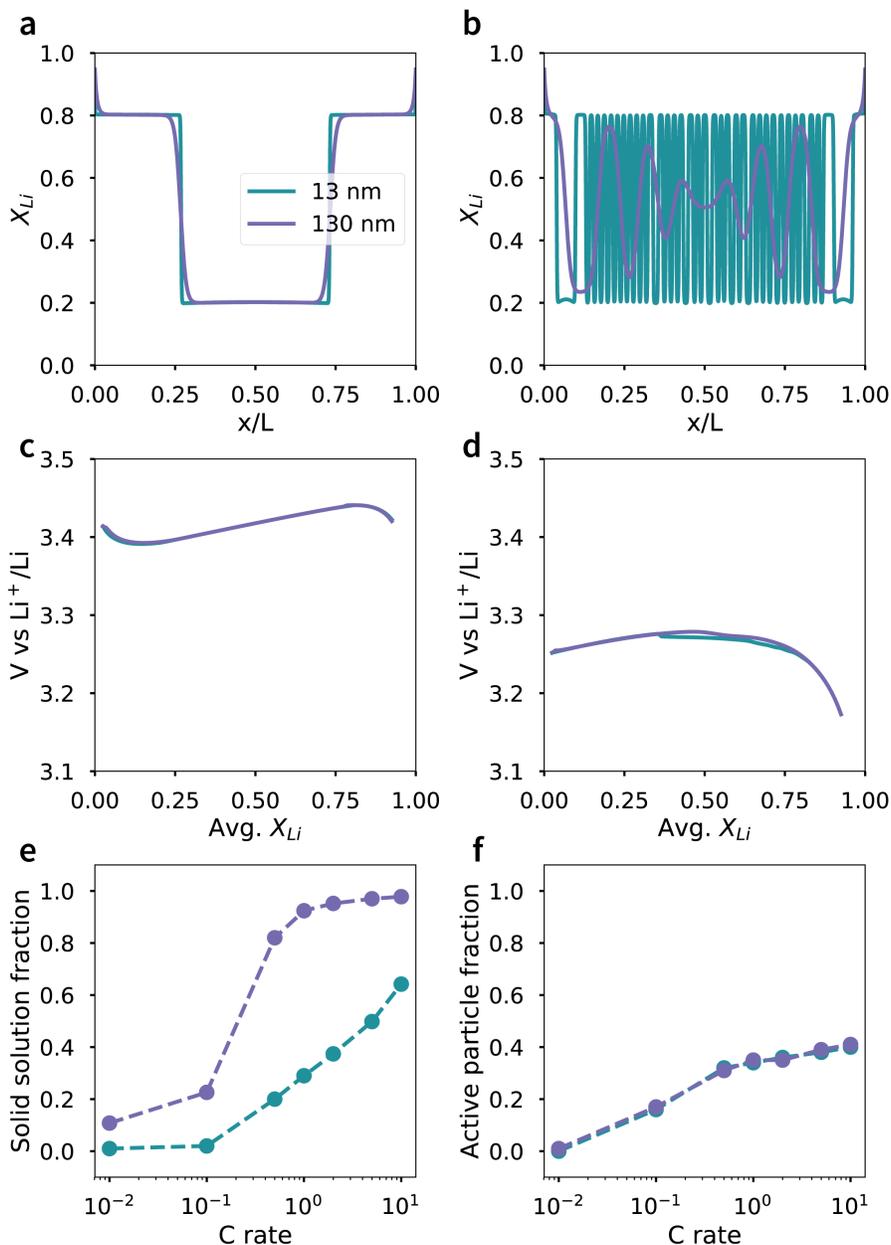

**Figure S13. 1D phase-field model sensitivity analysis on interface width. a,b**, 1D Li composition profile with different interface width at a discharge rate of 0.01C (a) and 2C (b). Spatial position x is normalized by the 1D particle size L. **c,d**, Voltage as a function of average Li composition. During a larger current discharge, the voltage is more sensitive to the interface width. **e**, The single particle solid solution fraction at 0.5 average Li during different constant current discharge. The critical current that suppresses intra-particle phase separation is changed by the interface width, the larger the width, the lower the critical current. **f**, Active particle fraction at 0.5 average Li during different constant current discharge. An ensemble of 50 particles of the same size with different a log-normal distribution in exchange current density were used to represent electrode level heterogeneity. Active particle fraction is insensitive to the interface width, as the particle size is significantly larger than the interface width in both scenarios.

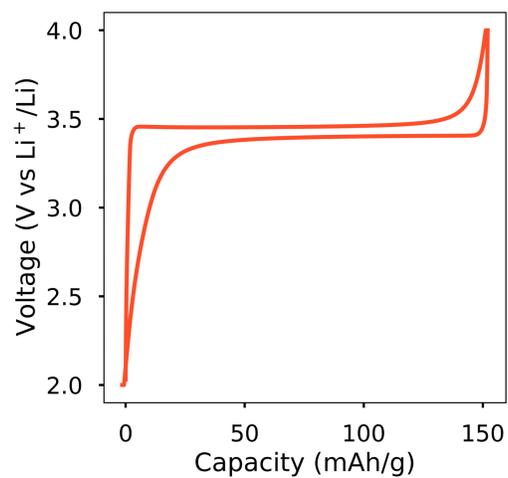

**Figure S14. Electrochemical cycling profile of platelet LFP.** The current is held at C/10.

# Supplementary Text

## I. Methods

### 1.1 Materials preparation and characterization

**Synthesis**

LiFePO$_4$ platelets were synthesized using a solvothermal method modified from previous work[2]. All the precursors were purchased from Sigma-Aldrich. In detail, 6 mL of 1 M H$_3$PO$_4$ was mixed with 24 mL of polyethylene glycol 400. Then, 18 mL of 1 M LiOH(aq) was added to precipitate Li$_3$PO$_4$. The mixture was constantly bubbled with dry N$_2$ at a flow rate of ~50 mL/min for ~16 h for deoxygenation. Next, 12 mL of deoxygenated H$_2$O was added via a Schlenk line to FeSO$_4$·7H$_2$O powder, which was pre-dried under vacuum. The FeSO$_4$ solution was then injected into the Li$_3$PO$_4$ suspension without oxygen exposure, and the entire mixture was transferred to a 100-mL Teflon-lined autoclave. The autoclave was heated to 140 ºC for 1 h and then to 210 ºC for 17 h. Note that this synthesis method produces smooth, flat, and well-faceted platelet particles but not void-free particles. The sample from this batched synthesis is referred to as BA1. The sample was annealed under Ar gas at 720 ºC for 5 h to eliminate anti-site defects[2–4]. Note that the sample for correlative imaging was not carbon coated.

**Delithiation of LiFePO$_4$**

To avoid interparticle phase separation, also known as a mosaic pattern[5], we chemically delithiated the pristine LiFePO$_4$ particles using a redox titration platform. This process ensured the formation of phase boundaries after phase separation and eliminated kinetic heterogeneities resulting from the carbon coating during electrochemical delithiation. Dilute 50 mL of H$_2$O$_2$(aq) was stoichiometrically added to 50 mL of the LiFePO4 particle suspension (1 mg/mL) to obtain Li$_{0.5}$FePO$_4$ (L$_{0.5}$FP) and FePO$_4$ (FP). To minimize morphological damage to the particles from the fast non-uniform local reaction, the titration rate was optimized to be 5 mL/h and the suspension was constantly stirred at 500 rpm.

**Electrochemical characterization**

Given the relevance of the particles for battery applications, we electrochemically tested the materials compared to commercial LFP powders. Specifically, pristine LiFePO$_4$ was carbon coated by mixing with sucrose (LFP:sucrose = 5:1) and annealing under Ar gas at 720 ºC for 5 h[2,6], cast as an electrode sheet, and then assembled into 2032-type lithium half cells to test the electrochemical cycling performance. The series of tests performed was 3 full C/10 charge and discharge cycles, followed by 3 C/5 full cycles, 3 1C full cycles, and eventually 2C full cycles. The voltage cutoff was set at 2 V and 4 V. All the cells were tested inside Arbin LBT20084 battery cyclers, held at 30 °C by an environment chamber. Electrodes made from BA1 achieved an average of 150 mAh/g capacity (Fig. S14). The tests confirmed the industrial relevance of the materials studied, particularly for battery applications.

## Structural and morphological characterization

For structural and morphological characterization, scanning electron microscopy (SEM), X-ray diffraction (XRD), and atomic force microscopy (AFM) analyses of the pristine and chemically delithiated samples were performed.

SEM was performed using a Sirion (FEI) machine at the Stanford Nano Shared Facility (SNSF). A primary diamond-shaped platelet structure with feature dimensions of 4 × 0.3 × 2 $\mu m^3$ (in crystallographic a,b,c directions) was observed for the BA1 particles. No significant morphological change was observed after chemical delithiation. We note that the BA1 particles contained morphological defects, which were later characterized using X-ray tomography.

XRD was first performed on a Bruker D8 at Stanford for structural characterization and at beamline 2-1 at the Stanford Synchrotron Radiation Lightsource (SSRL) at 17-keV beam energy. A LaB$_6$ standard reference material was used for calibration of the setup and instrumental broadening. The capillaries were 0.5-mm special glass capillaries (Charles Supper) and were loaded inside an Ar-filled glovebox. The samples were rotated continuously throughout the measurement to avoid preferential orientation effects. The beam was 0.5 mm in width and 1 mm in height. Raw data were recorded using a Pilatus 100K detector at a 700-mm distance from the capillary sample.

For AFM, the Li$_x$FePO$_4$ platelet was dispersed with isopropyl alcohol onto mica and imaged using an Asylum Cypher ES scanning probe microscope. Image processing including flattening was performed using the built-in Asylum software package based on Igor Pro.

### 1.2 Correlative X-ray microscopy and 4D-STEM

#### Sample dispersion for imaging

Pristine LiFePO$_4$, Li$_x$FePO$_4$, and FePO$_4$ platelets were dispersed with deionized water and sonicated for 2 h under an ice bath. The top layer was extracted, followed by the same procedure as the previous step until a clean, uniform particle suspension was formed. The particles were then dispersed onto a TEM grid (Ted Pella, 200 mesh with reference index). The TEM grid was loaded into a TEM (TitanX) at the National Center for Electron Microscopy and then at the COSMIC beamline at the Advanced Light Source for correlative imaging.

#### 4D-STEM

4D-STEM was performed at the National Center for Electron Microscopy using a TitanX, operated at an accelerating voltage of 300 kV with a convergence semi-angle of 0.48 mRAD using a 40-µm condenser (C2), a camera length of 600 mm. The probe size was 2.2–2.5 nm with a step size of 49.5 or 16.6 nm. During the imaging, a condensed electron beam was raster scanned across the sample, with a diffraction pattern acquired at each scan position. Local information on the lattice, including the strain and rotation, was obtained using py4DSTEM[7], an open-source python package available at https://github.com/py4dstem/py4DSTEM.

**X-ray microscopy**

Scanning transmission X-ray microscopy (STXM) measurements were performed at beamlines 11.0.2 and 7.0.1.2 (COSMIC) using a zone plate with outer zone width of 45 nm. Images at different X-ray energies were aligned using the ECC alignment algorithm[8]. The Fe $L_3$ edge was used to determine the local state of charge. $LiFePO_4$ and $FePO_4$ were selected as reference samples, with a full range energy stack taken at a step size of 0.2 eV from 696 to 740 eV. The local composition was thus determined based on the absorption at 5 selected energies (700 eV, 706.2 eV, 706.8 eV, 711.2 eV and 713.4 eV, covering 1 pre-edge, 2 rising edges, and 2 falling edges) using a non-negative least squares optimization algorithm. We deliberately avoided on-edge measurement because of the ill-measured photon statistics from thick samples.

Ptychography measurements were performed at beamline 7.0.1.2 (COSMIC) at ALS and were taken in double-exposure mode. Ptychographic reconstruction was performed using standard methods available in the SHARP-CAMERA software package with parallel computation (http://camera.lbl.gov/). The spatial resolution was calculated by Fourier ring correlation (FRC) to be 10 nm (½ bit threshold) at 710 eV (Fig. S3b). The same particle was imaged in STXM before and after the X-ray ptychography measurement, and no major changes in local composition were observed, suggesting low X-ray damage to the composition.

**Image registration**

4D-STEM and X-ray spectro-ptychography are used to probe the local lattice and chemistry of the sample and generate a high-dimensional data set from each set of measurements. To understand the correlation between the structure and chemistry, the two images were then registered through an affine transformation that maximizes the correlation between a lattice vector length (from 4D-STEM) to the composition (from X-ray spectro-microscopy using a customized MATLAB code. Using these features for the alignment produced results with very clear boundaries, where the simple "two-phase" model for $LiFePO_4$ and $FePO_4$ could be used to explain most of the observations.

**1.3 Tomography characterization**

Ptychography–tomography measurement was performed at beamline 7.0.1.2 (COSMIC) at ALS. The $Li_xFePO_4$ platelet was first imaged using 4D-STEM and then transferred onto a FIB lift-out grid (Ted Pella, Omniprobe) and imaged at 1300 eV. A series of 86 2D projections of the optical densities (ODs) were recorded over a wide angular range from –83° to +83°. Scattering contrast defined as the sum of squared OD and phase[9] was used for image registration. Finally, an OD voxel size of $17.3 \times 17.3 \times 17.3$ nm$^3$ was reconstructed using the sparse filtered reconstruction technique (SIRT) with 300 iterations, with a resolution of approximately 30 nm. The 3D tomogram reveals the existence of pre-existing voids inside the platelet particles.

## II. Inverse learning via PDE constrained optimization

**The forward problem statement**

We briefly review the forward model and then state the inverse problem and frame it as a PDE constrained optimization problem.

We first introduce the variables $X_{Li}$, $\mathbf{d}$, and $\epsilon$. Here, for $\mathbf{r} = (x; y) \in \mathbb{R}^2$, $X_{Li}(\mathbf{r}) \in [0,1]$ represents the lithium spatial distribution; $\mathbf{d}(\mathbf{r}) := (u(\mathbf{r}); v(\mathbf{r})) \in \mathbb{R}^2$ represents the displacement vector for each position; and $\epsilon$ represents the strain tensor, capturing the spatial derivative of $\mathbf{d}$, with the following relation:

$$\epsilon = \begin{bmatrix} \epsilon_{xx} & \epsilon_{xy} \\ \epsilon_{yx} & \epsilon_{yy} \end{bmatrix} = \frac{1}{2}(\nabla \mathbf{d} + \nabla \mathbf{d}^T) = \begin{bmatrix} \frac{\partial u}{\partial x} & \frac{1}{2}\frac{\partial u}{\partial y} + \frac{1}{2}\frac{\partial v}{\partial x} \\ \frac{1}{2}\frac{\partial u}{\partial y} + \frac{1}{2}\frac{\partial v}{\partial x} & \frac{\partial v}{\partial y} \end{bmatrix} \quad (2.1)$$

By definition, the strain tensor is symmetric and can be decomposed into two parts: the elastic strain and chemical strain (used interchangeably as compositional eigenstrain in our context):

$$\epsilon = \epsilon^{chem} + \epsilon^{el} \quad (2.2)$$

Each component follows a constitutive relation. The chemical strain is composition dependent, with a strain-composition relation denoted as $f$:

$$\epsilon^{chem} = \begin{bmatrix} \epsilon_{xx}^{chem} \\ \epsilon_{xy}^{chem} \\ \epsilon_{yy}^{chem} \end{bmatrix} = \begin{bmatrix} f_x(X_{Li}) \\ 0 \\ f_y(X_{Li}) \end{bmatrix} = \boldsymbol{f}(X_{Li}) \quad (2.3)$$

The elastic strain–stress relation is linear and can be parameterized by a stiffness tensor $\mathbf{C}(\alpha, \beta, \gamma, \omega)$ as the following, using the 2D plane stress condition (detailed in a later section) given the geometry of the particle. The $\mathbf{C}(\alpha, \beta, \gamma, \omega)$ tensor is known from the literature[10]. In vector form, it can expressed as

$$\begin{bmatrix} \sigma_{xx} \\ \sigma_{xy} \\ \sigma_{yy} \end{bmatrix} = \mathbf{C} \begin{bmatrix} \epsilon_{xx}^{el} \\ \epsilon_{xy}^{el} \\ \epsilon_{yy}^{el} \end{bmatrix} = \begin{bmatrix} \alpha & 0 & \beta \\ 0 & \omega & 0 \\ \beta & 0 & \gamma \end{bmatrix} \begin{bmatrix} \epsilon_{xx}^{el} \\ \epsilon_{xy}^{el} \\ \epsilon_{yy}^{el} \end{bmatrix} \quad (2.4)$$

Given this information, a forward model takes $X_{Li}, \boldsymbol{f}$, and $\mathbf{C}$ as input and calculates the displacement vector $\mathbf{d}$ by solving a strain-energy minimization problem: $\min_{\mathbf{d}} \frac{1}{2} \int \sigma \epsilon^{el} dS$, for which the Karush–Kuhn–Tucker solution is the stress-equilibrium condition, also known as the stress continuity equation, with a pure Neumann boundary condition ($\hat{\mathbf{n}}$ is the unit surface norm vector):

$$\nabla \cdot \sigma = 0, \quad \hat{\mathbf{n}} \cdot \sigma = 0 \quad (2.5)$$

Define $c := \begin{bmatrix} \alpha & 0 & 0 & \beta \\ 0 & \frac{\omega}{2} & \frac{\omega}{2} & 0 \\ 0 & \frac{\omega}{2} & \frac{\omega}{2} & 0 \\ \beta & 0 & 0 & \gamma \end{bmatrix}$, a forward model solves for **d** using the following equation:

$$\nabla \cdot (c \otimes \nabla \mathbf{d}) = \nabla \cdot (c \otimes f), \quad \hat{\mathbf{n}} \cdot (c \otimes \nabla \mathbf{d}) = \nabla \cdot (c \otimes f) \tag{2.6.1}$$

where the notation $c \otimes \nabla \mathbf{d}$ represents

$$c \otimes \nabla \mathbf{d} = \begin{bmatrix} \alpha & 0 & 0 & \beta \\ 0 & \frac{\omega}{2} & \frac{\omega}{2} & 0 \\ 0 & \frac{\omega}{2} & \frac{\omega}{2} & 0 \\ \beta & 0 & 0 & \gamma \end{bmatrix} \begin{pmatrix} \frac{\partial u}{\partial x} \\ \frac{\partial u}{\partial y} \\ \frac{\partial v}{\partial x} \\ \frac{\partial v}{\partial y} \end{pmatrix}$$

**The inverse learning problem**

With recent advancements in microscopy, we can now spatially resolve the lattice-change-related strain $\hat{\epsilon}$ and rotation $\hat{\theta}$, which can be converted into the displacement $\hat{\mathbf{d}}$ using an inverse gradient algorithm[11]. Given the composition $X_{Li}$ and $\hat{\mathbf{d}}$, our objective is to learn $f(X_{Li})$. Here, the hat notation is used to represent the experimental data. The inverse learning problem can therefore be abstracted into the following problem. For simplicity, we used the experimentally observed displacement as the boundary condition (rather than a pure Neumann boundary condition) to guarantee the uniqueness of the solution; nonetheless, similar steps can be taken to ensure uniqueness (such as Lagrangian formulations, see ref[12]) for pure Neumann boundary conditions.

*T1. Inverse composition–eigenstrain learning problem statement*

$$\begin{aligned} \min_{f} \quad & F := \| \hat{\mathbf{d}} - \mathbf{d} \|^2 \\ \text{s.t.} \quad & \nabla \cdot (c \otimes \nabla \mathbf{d}) = \nabla \cdot (c \otimes f) \\ & \mathbf{d}|_\Omega = \hat{\mathbf{d}}|_\Omega \end{aligned} \tag{2.7}$$

The finite element method (FEM) discretizes the PDE constraint into a finite number of algebraic equality constraints. We employed a 6-node triangular element to discretize the PDE constraint. Note that in the notation below, all the flattened column vectors are represented using a non-bold form. For example, for the vector $\mathbf{d} \in \mathbb{R}^2$ at multiple locations, its total nodal representation is flattened into $d \in \mathbb{R}^{2N_p}$, where $N_p$ is the number of nodes.

*T2. FEM-discretized inverse composition–eigenstrain learning problem statement*

$$\begin{aligned} &\min_{f} &&F := \| \hat{d} - d \|^2 \\ &\text{s.t.} &&K(\alpha, \beta, \gamma, \omega)d = G[\boldsymbol{f}] \end{aligned} \quad (2.8)$$

Here, $K(\alpha, \beta, \gamma, \omega) \succcurlyeq 0$ (semi positive definite) is the assembled stiffness matrix that is known, and $G$ is a column vector that is a linear transformation of the $f$ derivative values at each node.

We want to learn a function defined over $C[0,1]$. To ensure the well-posedness of the inverse problem, $\boldsymbol{f}$ may be represented by orthogonal basis functions. Here, we approximate $\boldsymbol{f}$ by a finite set of Legendre polynomial series with $f_x(X_{Li}) = \sum_{i=1}^{N} a_i L_i(2X_{Li} - 1)$, $f_y(X_{Li}) = \sum_{i=1}^{N} b_N L_n(2X_{Li} - 1)$ and keep $N$ as a hyper-parameter determined by data, to be discussed later. Using the method above, we can transform the right-hand side of the algebraic constraint into $G[f] = Dp$, with $D = [D_1, D_2, \ldots, D_N, D_1, \ldots, D_N]$, $D_i = G[L_n(X_{Li})]$ and $p := (a_1; a_2; \ldots; a_N; b_1; b_2; \ldots; b_N)$. Thus, the final optimization problem can be stated as the following:

*T3. Inverse composition–eigenstrain learning problem statement*

$$\begin{aligned} &\min_{p} &&F := \| \hat{d} - d \|^2 \\ &\text{s.t.} &&K(\alpha, \beta, \gamma, \omega)d = Dp \end{aligned} \quad (2.9)$$

The optimization problem is equivalent to a least-squares minimization problem, where the strong convexity nature ensures a unique optimal solution, which can be computed directly using matrix inversion. The confidence interval can also be evaluated for each entry in $p$ using basic maximum likelihood theory.

**Data-driven discovery of constitutive equations**

We first verified our inverse learning framework based on forward simulation results and then applied the same framework to the correlated image dataset. The learning result based on the simulation result is shown in Fig. S5a. The learning process should be robust against the approach of the discretization; we therefore deliberately picked the finite difference for the forward simulation and recovered the original constitutive relation using FEM discretization. For data-driven discovery of constitutive relations, we used images from phase-separated particles, as they are the only ones that contain information over the entire composition range of interest. The high-dimensional dataset, containing strain, rotation, and composition, was divided into test and training sets, as shown in Fig. S2. The training result with N as the hyperparameter is shown in Fig. S5b. The test image was then used to determine the appropriate N to avoid overfitting. This approach is similar to an L1 regularization that adds the sparsity-promoting term to the objective function, although setting the training this way allows for a more intuitive understanding of the structure of the functional space for the constitutive law. Eventually, we observed that N

= 2 produced the lowest test error, with any number beyond 2 resulting in a model overfit. The eigenstrain relation recovered is linear w.r.t. Li composition, for the most part. The result is consistent with many other single-phase solid-solution alloy systems that can be directly characterized[13].

## III. Phase-field simulation

### 2D phase-field simulation

The role of coherency (elastic) strain can be explored through a 2D phase-field simulation. We adopted a two-order parameter phase field to represent the particle thickness $\phi$ and concentration $X_{Li}$ (represented by $X$ for convenience), respectively, with the formulation below. Consistent with the notation in the inverse learning section, we first defined $\boldsymbol{d}$ and $\epsilon$ in a 2D space map with $\mathbf{r} = (x; y) \in \mathbb{R}^2$. $X(\boldsymbol{r}) \in [0,1]$ represents the lithium spatial distribution, and $\boldsymbol{d}(\boldsymbol{r}) := (u(\boldsymbol{r}); v(\boldsymbol{r})) \in \mathbb{R}^2$ represents the displacement vector for each position. For convenience, $x$ and $y$ are aligned with the crystallographic a and c directions, respectively. Parameterization of the phase-field model requires input of $\kappa$ and $\Omega$ for the concentration field $X$ and $U$ and $\xi$ for the thickness field $\phi$, defined below. All the parameters were tuned to accurately represent the system of scale, similar to existing literature[5,6,14,15]. Table S1 summarizes all the parameters used for the simulation. To form the diamond-shape geometry commonly observed in experiments, we adopted a parameterization of $\xi$ as a function of orientation, as shown in Fig. S7b. We note that the parameterization is not unique as long as all of the local minima maintain the same position and value to ensure the correct Wulff construction[16]. Specifically, in our case, parameterization is in the following form (see Table S1 for the coefficients):

$$\xi(\theta) = A_0 + \sum_{i=1}^{6} A_i \cos(B_i \theta) \tag{3.1}$$

The total Helmholtz energy is defined as the following:

$$H[\phi(\boldsymbol{r}), \boldsymbol{d}] = \int h^X + h^{el} + h^\phi \, dV \tag{3.2}$$

$$h^{X_{Li}} = \phi \Omega X_{Li}(1 - X_{Li}) + k_B T[X_{Li} \ln X_{Li} + (1 - X_{Li}) \ln(1 - X_{Li})] + \frac{1}{2}\kappa \phi |\nabla X_{Li}|^2 \tag{3.3}$$

$$h^{el} = \frac{1}{2}\phi \sum_{ij} \sigma_{ij} \epsilon_{ij}^{el} \tag{3.4}$$

$$h^\phi = U\phi^2(1-\phi)^2 + \xi(\hat{\boldsymbol{n}})^2 |\nabla \phi|^2 \tag{3.5}$$

We first initialized $\phi$ by fixing $X_{Li}$ at a constant value and then rapidly relaxed $\phi$ to form the correct particle shape. To speed up the convergence, a circular shape with the same area size as the desired particle was adopted. Conservation of $\phi$ was ensured using the Lagrangian method while minimizing $h^\phi$. After this, $\phi$ was fixed, and a random distribution of concentration $X$ was initialized. The concentration field was further relaxed driven by Cahn–Hilliard kinetics, which is a strictly diffusion-driven process. Of note, during each concentration update, chemo-mechanical equilibrium was reached (because elastic relaxation occurs much faster than ionic transport in our system) through the conjugate gradient method that minimizes the total elastic energy.

$$\frac{\partial X_{Li}}{\partial t} = \nabla \cdot \frac{D_0 X_{Li}(1-X_{Li})}{k_B T} \nabla \mu = \nabla \cdot \frac{D_0}{k_B T} X_{Li}(1-X_{Li}) \nabla \frac{\delta H}{\delta X_{Li}} \tag{3.6}$$

The finite difference method was implemented with periodic boundary conditions at vaccuum ($\phi = 0$). Formation of the particle is illustrated in Fig. S7a. The simulated relaxation process is complete after a numerical convergence is reached. The elastic strain field of an infinitely large particle as well as a platelet particle after phase separation are presented in Fig. S7c and d, where the chemo-mechanics are relaxed through coherency strain. The elastic strain field after phase separation, both in an infinitely sized particle as well as in a platelet-shaped particle show an almost negligible strain field across the simulation grid.

To generate additional chemo-mechanical insights, we removed the compositional eigenstrain and coherency strain from the experimental dataset. Here, we adopted 2D phase-field simulation to compute the coherency strain by initializing and freezing $\phi$ and $X$ to the experimental values and only allowed for the relaxation of stresses. The composition—eigenstrain relation from inversion was used in the process. The result is presented in Fig. S7e.

The residual strain field obtained from Fig. 3 and Fig. S9 shows significant heterogeneities. Although it is tempting to hypothesize that such heterogeneities are due to dislocations, pre-existing voids within particles may also produce a similar effect. Therefore, the non-uniformity of the thickness of the particle must be addressed. As the 2D model may fail to address the stress releases around pores, 3D phase-field simulation was adopted to understand the pore effects, as detailed in a later section.

Finally, we highlight the importance of the chemical expansion coefficient (differential) to the chemical potential:

$$\mu^{el} = \frac{\partial h^{el}}{\partial X_{Li}} = \frac{1}{2}\phi \frac{\partial}{\partial X_{Li}} \sum_{ij} \sigma_{ij}\left(\epsilon - \epsilon^{chem}(X_{Li})\right) = -\phi \sum_{ij} \sigma_{ij} \frac{\partial \epsilon_{ij}^{chem}}{\partial X_{Li}} \tag{3.7}$$

Because the chemical expansion coefficient directly contributes to the construction of the chemical potential, accurate characterization of the composition–eigenstrain relationship is thus highly relevant from a theoretical standpoint.

**Table S1. Parameters for 2D phase-field models.** R represents reverse calculated from theoretical geometry. M represents experimentally measured. I represents values inversely learned from experiments. The stiffness indices follow Voigt notations, i.e., {1, 2, 3, 4, 5, 6} map into {aa, bb, cc, bc, ac, ab}. The stiffness values are averaged from the FePO$_4$ and LiFePO$_4$ values. Note that for phase-field simulations, our analysis is insensitive to the values of diffusion constant, as our primary focus is on the final morphology.

| PARAMETERS | PHYSICAL MEANING | VALUES | UNITS | SOURCE |
|---|---|---|---|---|
| $k_B T$ | Thermal energy | 25.7 | $meV$ | |
| $c_{max}$ | Maximum concentration | 13.76 | $nm^{-3}$ | Ref[10,14] |

| Symbol | Description | Value | Unit | Source |
|---|---|---|---|---|
| $N_x$ | Simulation cell parameter | 300 | 1 | |
| $N_y$ | | 200 | 1 | |
| $h$ | | 3.3 | $nm$ | |
| $U$ | Shape interaction parameter | 205 | $meV$ | |
| $\xi$ | Surface anisotropy function | Fig. S7 | $eV \cdot nm^2$ | R |
| $M$ | Diffusion coefficient | 1e-14 | $\frac{cm^2}{s}$ | |
| $A_0$ | | 2.849 | | |
| $A_1$ | | -1.387 | | |
| $A_2$ | | -0.001 | | |
| $A_3$ | | -0.036 | $eV \cdot nm^2$ | |
| $A_4$ | | 0.061 | | |
| $A_5$ | | -0.004 | | |
| $A_6$ | Surface anisotropy function parameter | -0.036 | | R |
| $B_1$ | | 0.452 | | |
| $B_2$ | | 49.300 | | |
| $B_3$ | | 10.040 | $\frac{1}{rad}$ | |
| $B_4$ | | 5.672 | | |
| $B_5$ | | 7.190 | | |
| $B_6$ | | 4.502 | | |
| $\Omega$ | Composition interaction parameter | 113.08 | $meV$ | Ref[6,14,15] |
| $\kappa$ | Phase-field gradient penalty parameter | 13.4 | $eV \cdot nm^2$ | M |
| $D_0$ | Lithium diffusivity | 1e-14 | $\frac{cm^2}{s}$ | Ref[14,17] |
| C11 | | 157.4 | | |
| C22 | | 175.8 | | |
| C33 | | 154 | | |
| C44 | | 37.8 | | |
| C55 | Elastic stiffness | 49.05 | $GPa$ | Ref[10,14,15] |
| C66 | | 51.6 | | |
| C12 | | 51.2 | | |
| C13 | | 53.25 | | |
| C23 | | 32.7 | | |
| $\epsilon_{aa}^0$ | Compositional eigenstrain | Fig. 2 | 1 | I |

| $\epsilon_{cc}^0$ | | 1 | I |
|---|---|---|---|

**Plane-stress model**

For faster computation, the mechanics and chemistry of the particle can be mostly captured well with a 2D plane-stress model given the platelet shape of the particle. Assuming direction 2 is the depth direction, we derive the plane-stress condition:

$$\begin{aligned}
\sigma_{11} &= C_{1111}\epsilon_{11} + C_{1122}\epsilon_{22} + C_{1133}\epsilon_{33} \\
\sigma_{22} &= C_{1122}\epsilon_{11} + C_{2222}\epsilon_{22} + C_{2233}\epsilon_{33} \\
\sigma_{33} &= C_{1133}\epsilon_{11} + C_{1133}\epsilon_{22} + C_{3333}\epsilon_{33} \\
\sigma_{13} &= C_{1313}\epsilon_{13} \\
\sigma_{12} &= \sigma_{23} = \sigma_{22} = 0
\end{aligned} \quad (3.8)$$

Upon comparison with our final equation, it is apparent that

$$\begin{aligned}
\alpha &= C_{1111} - \frac{C_{1122}^2}{C_{2222}} \\
\beta &= C_{1133} - \frac{C_{1122}C_{2233}}{C_{2222}} \\
\gamma &= C_{3333} - \frac{C_{2233}^2}{C_{2222}} \\
\omega &= C_{1313}
\end{aligned} \quad (3.9)$$

For brevity, the stiffness tensor will be represented using Voigt notation, i.e., {1, 2, 3, 4, 5, 6} map into {aa, bb, cc, bc, ac, ab}, as shown in Table S1 and Table S4.

**3D phase-field simulation and void effect**

To understand the void effect on phase separation and the elastic strain field, 3D phase-field simulations were performed on a single $Li_{0.5}FePO_4$ particle with dimensions of 500 nm × 130 nm × 100 nm. Our goal is to investigate whether voids, as observed from tomography experiment, may generate significant strain heterogeneities in the particle. The eigenstrain in the crystallographic b direction (the depth direction in platelet particles) is assumed to be linear w.r.t. Li composition, with a constant chemical expansion coefficient of 0.0346, consistent with previous experimental measurements[18] and computational studies[15,19,20]. The gradient energy penalty in the b direction, $\kappa_b$, is assumed to be 100 times that of the in-plane direction $(\kappa_b = 100\kappa_{a/c})$. This particular value for $\kappa_b$ was previously found essential for establishing the 2D depth-averaged model under fast lithiation conditions[14,15,19]. For relaxation, the diffusivity in the b direction is taken to be $10^{-9}$ cm$^2$s$^{-1}$, consistent with first-principles calculations[21]. All the other parameters are consistent with those in Table S1. The particle is subject to zero external force and the strain fields are considered to be coherent. The set of differential equations of our model was discretized using finite elements, and all unknowns were approximated with linear polynomial basis functions[22]. For the time integration of the resulting system of differential algebraic equations, we used 2nd order Gear method.

Two scenarios were explored: the void at the center of the particle and the void near the edge of the particle. The radius of the void is 40 nm, similar to that of the tomography experiment. After relaxation, the interface intersects the void in the first case, whereas in the second, the void is completely embedded in one phase. For better comparison with the experimental result, the b-direction averaged concentration and strain values were calculated. As shown in Fig. S8, negligible elastic strain at the scale-bar level (< 0.6%) was generated in a coherent interface model, hence indicating that the experimentally observed strain hotspots are likely due to other factors such as dislocations.

Finally, we performed relaxation simulations with isotropic gradient energy penalty along all directions. We found that our results are identical to the anisotropic case, as interfaces cannot form in the b direction for particles of this specific thickness (~100 nm). The reason behind this result lies on the anisotropic elastic nature of LiFePO$_4$, where it has been previously shown[20] that interfaces perpendicular to the b direction require excess elastic energy to be stored in the system.

### 1D phase-field simulation and model sensitivity to interface width

1D phase-field simulation was performed to explore the phase transformation pattern, voltage, solid solution and active particle fraction sensitivity to interface width. By varying the $\kappa$, and fixing the rest of the parameters, we were able to explore scenarios when the interface is 130 nm and 13 nm for a 1.3 μm particle, as the interface width $\sim\sqrt{\kappa}$. The result is shown in Fig. S13. We briefly list all relevant equations below:

$$\frac{\partial X^i}{\partial t} = -\nabla \cdot J_D^i + R^i + \xi^i \tag{3.10}$$

$$\mu_{Li}^i = \Omega(1 - 2X) + k_B T \ln\left(\frac{X^i}{1 - X^i}\right) - \frac{\kappa}{\rho_s}\nabla^2 X + \frac{B}{\rho_s}(X - \bar{X}) \tag{3.11}$$

$$R^i(X^i, \Delta\phi, \mu_{Li}^i) = k_0^i X^{i\frac{1}{2}}(1 - X^i)^{\frac{3}{2}} sinh\left(\frac{e(\Delta\phi + \mu_{Li}^i)}{2k_B T}\right) \tag{3.12}$$

$$J_D^i(X^i, \mu_{Li}^i) = -\frac{D_0}{k_B T} X^i(1 - X^i)\nabla\mu_{Li}^i \tag{3.13}$$

$$X^i\big|_{x=0,L} = c_s \tag{3.14}$$

$$\nabla\mu^i\big|_{x=0,L} = 0 \tag{3.15}$$

$$I = \sum_i I^i = e\rho_s \sum_i \int_0^L \frac{\partial X^i}{\partial t} dx \tag{3.16}$$

$X$ is the spatial lithium composition, $t$ the time, $R$ the reaction current, $\mu$ the lithium variational chemical potential, $J_D$ is the effective surface diffusional current, $\xi$ the Langevin thermal noise, and index $i$ the $i^{th}$ particle in the simulation. $B$ is the coherency stress coefficient, $\kappa$ the chemical interfacial gradient penalty, $k_B$ the Boltzman constant, and $T$ the temperature ($300\ K$ in our case). The model is consistent with previous literature[14,15,19]. Most values were taken from Table S1, except for the $\kappa$ corresponding to 130 nm thickness, which is taken as 100 times the original value, and $B$ (0.19 GPa) and $k_0$ (0.4 h$^{-1}$). Additionally, for the multi-particle simulation, the exchange current density prefactor $k_0^i$ follows a log-normal distribution: $\log(k_0^i/k_0) \sim N(0,1)$.

## IV. X-ray line profile analysis and identification of dislocation systems

The local hotspots in the residual strain map of $Li_xFePO_4$ indicate the presence of dislocations. To validate this conclusion, we performed fine structure analysis of synchrotron X-ray diffraction peaks to evaluate the dislocation Burgers vectors and densities inside platelet particles. Williamson–Hall analysis[23] has conventionally been employed to analyze and differentiate the crystallite-size-induced and strain-induced broadening effect; however, the anisotropy of the sample makes the data interpretation challenging[24]. To gain deeper insight into the dislocation distribution and density, the variance method[25,26] was used to analyze the dislocations. We present the relevant equations below and refer readers to references[25,26] for detailed documentation.

Let q be the reciprocal space representation defined as
$$q = \frac{2}{\lambda}(\sin(\theta) - \sin(\theta_0)) \tag{4.1}$$
where $\lambda$ is the wavelength and $\theta$ is the scattering angle, with $\theta_0$ representing the scattering angle at the center of the relevant peak of interest. The k-th order restricted moment of the intensity distribution is defined as
$$M_k(q) = \frac{\int_{-q}^{q} r^k I(r) dr}{\int_{-\infty}^{+\infty} I(r) dr} \tag{4.2}$$

The size and dislocation density can be estimated using the following equation:

$$M_2(q) = \frac{1}{\pi^2 \epsilon_F} q - \frac{L}{4\pi^2 K^2 \epsilon_F^2} + \frac{\rho^*}{2\pi^2} \ln\left(\frac{q}{q_0}\right) \tag{4.3}$$

$$\frac{M_4(q)}{q^2} = \frac{1}{3\pi^2 \epsilon_F} q + \frac{\rho^*}{4\pi^2} + \frac{3 <\rho^{*2}>}{4\pi^2 q^2} \ln^2\left(\frac{q}{q_1}\right) \tag{4.4}$$

Here, $\epsilon_F$ is the average column size (particle size), $K$ is the Scherrer constant, $L$ is the taper parameter, and $\rho^*$ is the effective dislocation density and can be expressed as $\rho^* = \sum_i^N \Lambda_i \rho_i$. Here, $\rho_i$ is the density of a given dislocation type, $\Lambda_i$ is a geometric factor, and $N$ is the total number of candidate dislocations. The geometric factor can be further expressed as $\Lambda_i = \frac{\pi}{2} g^2 b_i^2 C_i$, where $g$ is the magnitude of the diffraction vector and $b_i^2$ is the square of the Burgers vector. The dislocation contrast factor $C_i$ depends on the Burgers vector, elastic stiffness of the material, diffraction vector, and dislocation line vector. $<\cdot>$ is the operator for evaluating the mean, and $q_0$ and $q_1$ are two fitting parameters for normalization in the logarithmic term. As shown in the equations above, for M4 and M2, the size and dislocation broadening effect are deconvoluted. We thus use the two equations to analyze the effective defect density $\rho^*$. Of note, the M2 and M4 results were consistent with each other, serving as an internal benchmark in our analysis.

Fig. 4 summarizes the effective dislocation density of 5 diffraction peaks (101, 210, 111, 301, 311) determined experimentally. In general, a higher effective density can be observed in phase-separated particles and fully delithiated particles at most angles. Theoretically, each dislocation system carries a unique angle-dependent contrast factor, which is a function of its Burgers vector, elastic stiffness, and crystal structure. Decomposition of the effective experimental dislocation density into possible dislocation candidates can therefore generate insight into the activated dislocation structure. Based on existing literature on LiFePO$_4$, other olivine systems[27–29], and the preferred dislocation line direction due to the platelet geometry, we considered 26 candidate dislocation systems, shown in Table S3. Edge and mixed dislocations were considered due to the mismatch strain relief during phase separation. We first screened the candidates by selecting the dislocations with relatively low dislocation energy, estimated based on $Kb^2$, where $K$ is the dislocation energy factor and $b$ is the Burgers vector,[30]. resulting in 15 slip systems as potential candidates to understand the diffraction peak broadening. The inputs for contrast factors are provided in Table S4, and the calculated values are presented in Table S2.

**Table S2. Dislocation contrast factors.** Edge (E) and mixed (M) dislocations are listed.

| | # | Slip system | $l$ | Type | 101 | 210 | 111 | 301 | 311 |
|---|---|---|---|---|---|---|---|---|---|
| FP | 1 | ½[0 1 1](0 $\bar{1}$ 1) [1 0 0] | | E | 0.1623 | 0.0529 | 0.3046 | 0.0252 | 0.0764 |
| | 2 | ½[1 1 0]($\bar{1}$ 1 0) [0 0 1] | | E | 0.0151 | 0.3812 | 0.0701 | 0.1892 | 0.2184 |
| | 3 | [0 0 1](1 0 0) [0 1 0] | | E | 0.4846 | 0.0307 | 0.2012 | 0.2459 | 0.1659 |
| | 4 | [0 1 0](0 0 1) [1 0 0] | | E | 0.0488 | 0.0830 | 0.2847 | 0.0076 | 0.0718 |
| | 5 | [0 1 0](1 0 0) [0 0 1] | | E | 0.0029 | 0.3187 | 0.1073 | 0.0360 | 0.1222 |
| | 6 | [0 0 1](0 1 0) [1 0 0] | | E | 0.3311 | 0.0108 | 0.3368 | 0.0514 | 0.0849 |
| | 7 | ½[0 1 1](1 0 0) [0 1 $\bar{1}$] | | M | 0.1333 | 0.1415 | 0.2614 | 0.1072 | 0.1556 |
| | 8 | ½[1 1 0](0 0 1) [1 $\bar{1}$ 0] | | M | 0.1783 | 0.0650 | 0.2820 | 0.1072 | 0.0927 |
| | 9 | ½[1 0 1](0 1 0) [$\bar{1}$ 0 1] | | M | 0.1709 | 0.2093 | 0.1587 | 0.2390 | 0.2411 |
| | 10 | [0 1 0](1 0 0) [0 1 1] | | M | 0.0298 | 0.2599 | 0.1407 | 0.0361 | 0.1267 |
| | 11 | [0 1 0](1 0 0) [0 2 1] | | M | 0.0193 | 0.2307 | 0.1961 | 0.0175 | 0.1327 |
| | 12 | ½[0 1 1](1 0 0) [0 1 0] | | M | 0.1967 | 0.1332 | 0.2352 | 0.0998 | 0.1502 |
| | 13 | [0 1 0](0 0 1) [1 1 0] | | M | 0.0476 | 0.0866 | 0.2694 | 0.0196 | 0.0814 |
| | 14 | [0 1 0](0 0 1) [1 2 0] | | M | 0.0338 | 0.1070 | 0.2602 | 0.0245 | 0.0982 |
| | 15 | ½[0 1 1](0 0 1) [0 1 0] | | M | 0.1670 | 0.1799 | 0.1361 | 0.3077 | 0.2437 |
| LFP | 1 | ½[0 1 1](0 $\bar{1}$ 1) [1 0 0] | | E | 0.1471 | 0.0516 | 0.3416 | 0.0263 | 0.0942 |
| | 2 | ½[1 1 0]($\bar{1}$ 1 0) [0 0 1] | | E | 0.0143 | 0.3512 | 0.0533 | 0.2069 | 0.2057 |
| | 3 | [0 0 1](1 0 0) [0 1 0] | | E | 0.4645 | 0.0430 | 0.2040 | 0.2826 | 0.1914 |
| | 4 | [0 1 0](0 0 1) [1 0 0] | | E | 0.0464 | 0.0772 | 0.3196 | 0.0083 | 0.1914 |
| | 5 | [0 1 0](1 0 0) [0 0 1] | | E | 0.0052 | 0.3240 | 0.0862 | 0.0756 | 0.1342 |
| | 6 | [0 0 1](0 1 0) [1 0 0] | | E | 0.3111 | 0.0101 | 0.3776 | 0.0556 | 0.1041 |
| | 7 | ½[0 1 1](1 0 0) [0 1 $\bar{1}$] | | M | 0.1167 | 0.1699 | 0.2389 | 0.1166 | 0.1637 |
| | 8 | ½[1 1 0](0 0 1) [1 $\bar{1}$ 0] | | M | 0.1611 | 0.0639 | 0.3081 | 0.0553 | 0.1090 |
| | 9 | ½[1 0 1](0 1 0) [$\bar{1}$ 0 1] | | M | 0.1543 | 0.2146 | 0.1537 | 0.2366 | 0.2423 |

| | | | | | | | | |
|---|---|---|---|---|---|---|---|---|
| 10 | [0 1 0](1 0 0) | [0 1 1] | M | 0.0246 | 0.2754 | 0.1339 | 0.0489 | 0.1372 |
| 11 | [0 1 0](1 0 0) | [0 2 1] | M | 0.0137 | 0.2464 | 0.1865 | 0.0202 | 0.1380 |
| 12 | ½[0 1 1](1 0 0) | [0 1 0] | M | 0.1767 | 0.1497 | 0.2290 | 0.1075 | 0.1604 |
| 13 | [0 1 0](0 0 1) | [1 1 0] | M | 0.0430 | 0.0822 | 0.2935 | 0.0218 | 0.0949 |
| 14 | [0 1 0](0 0 1) | [1 2 0] | M | 0.0303 | 0.0999 | 0.2660 | 0.0298 | 0.1044 |
| 15 | ½[0 1 1](0 0 1) | [0 1 0] | M | 0.1293 | 0.1973 | 0.1181 | 0.3023 | 0.2402 |

Table S3. Potential dislocation types

| Glide plane, n | Label | Burgers vector, b | Line vector, l | Dislocation type | $K$ [GPa] | $Kb^2$ [nJ/m] | Source |
|---|---|---|---|---|---|---|---|
| (100) |   | [010] | [010] | screw | 41.85 | 15.3 | Ref [27,28] |
|   | A | [010] | [001] | edge | 73.75 | 27.0 |   |
|   | B | [010] | [011] | mixed | 59.11 | 21.6 |   |
|   | C | [010] | [021] | mixed | 51.79 | 18.9 |   |
|   |   | [001] | [001] | screw | 43.15 | 9.7 |   |
|   | D | [001] | [010] | edge | 72.18 | 16.2 |   |
|   |   | ½[011] | [011] | screw | 57.18 | 8.4 |   |
|   | E | ½[011] | [01$\bar{1}$] | mixed | 63.49 | 9.4 |   |
|   | F | ½[011] | [010] | mixed | 56.68 | 8.4 |   |
| (010) |   | [001] | [001] | screw | 43.15 | 9.7 | Ref [27] |
|   | G | [001] | [100] | edge | 67.32 | 15.1 |   |
|   |   | [100] | [100] | screw | 49.08 | 53.6 |   |
|   |   | [100] | [001] | edge | 61.77 | 67.5 |   |
|   |   | ½[101] | [101] | screw | 47.90 | 15.8 |   |
|   | H | ½[101] | [$\bar{1}$01] | mixed | 58.06 | 19.1 |   |
| (001) |   | [010] | [010] | screw | 41.85 | 15.3 |   |
|   | I | [010] | [100] | edge | 72.02 | 26.4 |   |
|   | J | [010] | [110] | mixed | 58.90 | 21.6 |   |
|   | K | [010] | [120] | mixed | 53.83 | 19.7 |   |
|   |   | [100] | [100] | screw | 49.08 | 53.6 |   |
|   |   | [100] | [010] | edge | 64.67 | 70.6 |   |
|   |   | ½[110] | [110] | screw | 46.77 | 17.0 |   |
|   | L | ½[110] | [1$\bar{1}$0] | mixed | 63.30 | 23.1 |   |
|   | M | ½[110] | [010] | mixed | 66.16 | 24.1 |   |
| (0$\bar{1}$1) | N | ½[011] | [100] | edge | 70.23 | 10.4 | Ref [29] |
| ($\bar{1}$10) | O | ½[110] | [001] | edge | 64.78 | 23.6 |   |

A non-negative least squares minimization was adopted to identify the major dislocation types, and the results are summarized in Table 1 and Fig. S11. For LFP and FP, the problem is defined as the following:

$$\min_{x} \quad F := \| \rho^* - \Lambda\rho \|^2$$
$$\text{s.t.} \quad \rho \geq 0$$
(4.5)

Here, $\rho^* \in \mathbf{R}^{5\times 1}$ is the column vector representing the effective dislocation density at 5 reflection angles. $\Lambda \in \mathbf{R}^{5\times 15}$ is the coefficient matrix, with each column representing the corresponding geometric factors for each slip system at 5 reflection angles. $\rho \in \mathbf{R}^{15\times 1}$ is the density for each type of dislocation. As observed in the setup, the minimization problem is convex, and its optimal solution can be recovered. The uniqueness of the solution was then verified by setting the nonzero values in $x$ to be zero, and a larger minimum value was reached after the same convex optimization.

We performed two types of analysis for FP: (1) dislocation types that exist in L50FP and LFP and (2) all candidates. The results are presented in Table S5. When the results were constrained to the 5 candidates that have been observed in LFP and $L_X$FP, 85% of the data can be explained by 4 dislocation types (labeled by FP). However, when all 15 candidates were used, our min-norm solution can explain 92% of the data (labeled by FP*); however, the dislocation type significantly deviated from LFP and $L_X$FP. Given that the variance method has roughly 15% error[26], which we set as our tolerance bound, we believe that the first case is more likely. Additionally, sensitivity analysis was performed for type 'D' and 'I', which showed that a value of 100 µm$^{-2}$ of 'D' causes an additional 2% error beyond the tolerance bound, whereas a value of 100 of 'I' introduced an additional ~1% error. Therefore, we believe the presence of 'D' in FP based on diffraction is unlikely, whereas 'I' may be underestimated. Nonetheless, the overall dislocation density is higher. Importantly, the first case is also consistent with residual strain analysis.

**Table S4. Lattice parameters (Å) and elastic stiffness constants (GPa) used as input for contrast factor calculation.** The elastic stiffness constants are indexed following Voigt notation.

|  |  | FePO$_4$ | LiFePO$_4$ |
|---|---|---|---|
| Lattice parameters (Å) | a | 9.96 | 10.45 |
|  | b | 5.88 | 6.05 |
|  | c | 4.86 | 4.74 |
| Elastic stiffness constants (GPa) | c11 | 175.9 | 138.9 |
|  | c22 | 153.6 | 198.0 |
|  | c33 | 135.0 | 173.0 |
|  | c44 | 38.8 | 36.8 |
|  | c55 | 47.5 | 50.6 |
|  | c66 | 55.6 | 47.6 |
|  | c12 | 29.6 | 72.8 |

|  |  | FePO$_4$ | LiFePO$_4$ |
|---|---|---|---|
|  | c13 | 54.0 | 52.5 |
|  | c23 | 19.6 | 45.8 |

**Table S5. FP analysis**

| n | Label | b | l | Type | FP | FP* |
|---|---|---|---|---|---|---|
|  | **Total density (μm$^{-2}$)** |  |  |  | 4619(351) | 4280(281) |
|  | **Maximum data explained %** |  |  |  | 85 | 92 |
| (100) | A | [010] | [001] | Edge | 1298(183) |  |
|  | B | [010] | [011] | Mixed |  |  |
|  | C | [010] | [021] | Mixed |  |  |
|  | D | [001] | [010] | Edge |  |  |
|  | E | ½[011] | [01$\bar{1}$] | Mixed |  |  |
|  | F | ½[011] | [010] | Mixed |  | 2726(274) |
| (010) | G | [001] | [100] | Edge |  |  |
|  | H | ½[101] | [$\bar{1}$01] | Mixed | 1382(146) |  |
| (001) | I | [010] | [100] | Edge |  |  |
|  | J | [010] | [110] | Mixed | 1939(261) |  |
|  | K | [010] | [120] | Mixed |  |  |
|  | L | ½[110] | [1$\bar{1}$0] | Mixed |  |  |
|  | M | ½[110] | [010] | Mixed |  | 1388(253) |
| (0$\bar{1}$1) | N | ½[011] | [100] | Edge |  |  |
| ($\bar{1}$10) | O | ½[110] | [001] | Edge |  | 166(131) |

Overall, Table 1 shows that there are 3–4 major slip systems that contribute to ~85% of the broadening effect. The overall dislocation density increased with the extent of delithiation, indicating the generation of dislocations as delithiation continues. Compared with LiFePO$_4$, phase-separated particles contain a significantly higher density in type 'D', which we identify as the most likely candidate causing the local hotspots and residual strain incompatibility observed in Fig. 3.

## V. Residual strain and compatibility

The residual strain is calculated by subtracting the chemical strain (compositional eigenstrain) and coherency (elastic) strain from the total strain measured by 4D-STEM:

$$\epsilon^{res} = \epsilon - (\epsilon^{chem} + \epsilon^{el}_{coherency}) \tag{5.1}$$

where the coherency strain is computed by 2D phase-field simulation, with zero dislocation density.

As we attribute the strain heterogeneities to the strain field of dislocations, we quantify the heterogeneities by calculating the residual of the strain compatibility equation:

$$r := \frac{\partial^2 \epsilon^{res}_{cc}}{\partial a^2} + \frac{\partial^2 \epsilon^{res}_{cc}}{\partial a^2} - 2\frac{\partial^2 \epsilon^{res}_{ac}}{\partial a \partial c} \tag{5.2}$$

where a and c represent the crystallographic directions. The mean squared residual per particle is shown in Fig. S9. We observe a peak value for particles at intermediate Li compositions, indicating the potential dislocation evolution during delithiation. This trend is consistent with the XRD analysis, where the dislocation (type 'D') that contributes to the ac plane reaches a maximum for intermediate Li composition particles ($L_X$FP).

## VI. Dislocation-density optimization

We considered an additional dislocation-induced elastic strain field in our model to further improve the model accuracy, specifically for the correlative dataset in Fig. 3. Based on linear elasticity, the total lattice strain is a superposition of individual components. We therefore modify the inverse model in the previous section to obtain

$$\epsilon = \epsilon^{chem} + \epsilon^{el}_{coherency} + \epsilon^{el}_{dis} \tag{6.1}$$

Again, $\epsilon$ is the total strain measurable by 4D-STEM, $\epsilon^{el}_{coherency}$ is the coherent elastic strain counterpart, and $\epsilon^{el}_{dis}$ is the elastic strain field induced by a dislocation. The elastic strain is subject to the stress-equilibrium condition:

$$\nabla \cdot (C : \epsilon^{el}_{dis}) = 0 \tag{6.2}$$
$$\nabla \cdot (C : \epsilon^{el}_{coherency}) = 0 \tag{6.3}$$

The single-dislocation elastic strain field can be computed using the Willis–Steeds–Lothe formula[31], with a given Burgers vector and line direction. Parameterizing the dislocation density field as $\rho$, the first equilibrium condition can be solved as

$$\epsilon^{el}_{dis}(r_0) = \int \rho(r) \epsilon^d(r_0 - r) d\Omega \tag{6.4}$$

Here, $\epsilon^d(r_0 - r)$ is the single dislocation field at position $r_0$ with a core located at r.

Dislocation-density optimization can thus be performed by enforcing the second stress equilibrium condition:

$$\begin{aligned} \min_{f} \quad & F := \|\hat{d} - d\|^2 \\ \text{s.t.} \quad & \nabla \cdot \left(c \otimes (\nabla \mathbf{d} - f - \epsilon^{el}_{dis})\right) = 0, \\ & \hat{n} \cdot \left(c \otimes (\epsilon - f - \epsilon^{el}_{dis})\right) = 0. \end{aligned} \tag{6.5}$$

The problem can be transformed into the following:

$$\begin{aligned} \min_{p} \quad & F := \|\hat{d} - d\|^2 \\ \text{s.t.} \quad & K(\alpha, \beta, \gamma, \omega) d = Dp + L\rho \end{aligned} \tag{6.6}$$

where D is the matrix whose column vector is the assembled Legendre polynomial and L is the discretization matrix, where each column vector represents the divergence of the dislocation stress field for each source at the corresponding location. Mathematically, the problem is equivalent to M1 and can be solved similarly. We present the result in Fig. S6. Sequential least squares regression was used to obtain the final result.

## VII. Discussion on model error

Here, we briefly compare the three models mentioned in the main text. Using a statistical regression framework, we can obtain the following formulation for the total strain. For consistency, $\epsilon$ is the total strain measured by 4D-STEM.

The baseline model (M0) from direct correlation ignores mechanical strain interaction; therefore,

$$M0: \epsilon = \epsilon^{chem}(X_{Li}) + s_0 \tag{7.1}$$

where $s_0 \sim N(0, \sigma_0^2)$ is the Gaussian noise.

The coherency strain model (M1) considers chemical eigenstrain and elastic strain, with the following:

$$M1: \epsilon = \epsilon^{chem}(X_{Li}) + \epsilon^{el}_{coherency} + s_1 \tag{7.2}$$

where $s_1 \sim N(0, \sigma_1^2)$ is the Gaussian noise.

Finally, the dislocation model considers elastic strain (composed of coherency and dislocation strain) and compositional eigenstrain that's measurable by 4D-STEM:

$$M2: \epsilon = \epsilon^{chem}(X_{Li}) + \epsilon^{el}_{coherency} + \epsilon^{el}_{dis} + s_2 \tag{7.3}$$

where $s_2 \sim N(0, \sigma_2^2)$ is the Gaussian noise.

Because setting $\epsilon^{el}_{coherency}$ and $\epsilon^{el}_{dis}$ to zero reduces M1 and M2 to M0, we have $\sigma_2^2 \leq \sigma_1^2 \leq \sigma_0^2$, which is confirmed by data, shown in Fig. 3e.